\documentclass[acmsmall]{acmart}
\usepackage{booktabs}

\usepackage{tabularx}
\usepackage{array}
\usepackage{amsmath}
\usepackage{ragged2e}

\newcommand{\thinmidrule}{\specialrule{0.03em}{\aboverulesep}{\belowrulesep}}

\newcolumntype{L}[1]{>{\RaggedRight\arraybackslash\hyphenpenalty=10000\exhyphenpenalty=10000\relax}p{#1}}

\newcolumntype{Y}{>{\RaggedRight\arraybackslash\hyphenpenalty=10000\exhyphenpenalty=10000\relax}X}
\AtBeginDocument{%
  }

\acmDOI{XXXXXXX.XXXXXXX}

\acmJournal{CSUR}

\begin{document}

\title{Security and Privacy in Large-Model-Driven Embodied Agents: Attacks, Defenses, and Future Directions}

\author{Lele Zheng}
\email{zhenglele@xidian.edu.cn}

\affiliation{%
  \institution{Xidian University}
  \city{Xi'an}
  \country{China}
}

\author{Tong Chen}
\affiliation{%
  \institution{Xidian University}
    \city{Xi'an}
  \country{China}}
\email{25031212204@stu.xidian.edu.cn}

\author{Ke Cheng}
\affiliation{%
  \institution{Xidian University}
  \city{Xi'an}
  \country{China}
}
\email{chengke@xidian.edu.cn}

\author{Tao Zhang}
\affiliation{%
 \institution{Xidian University}
   \city{Xi'an}
 \country{China}}

\author{Xingchi Liu}
\affiliation{%
  \institution{Xidian University}
    \city{Xi'an}
  \country{China}}

\author{Ji He}
\affiliation{%
  \institution{Xidian University}
  \city{Xi'an}
  \country{China}}
\email{jihe@xidian.edu.cn}

\author{Xutong Mu}
\affiliation{%
  \institution{Xidian University}
   \city{Xi'an}
  \country{China}}
\email{muxutong@xidian.edu.cn}

\author{Yulong Shen}
\affiliation{%
  \institution{Xidian University}
   \city{Xi'an}
  \country{China}}
\email{ylshen@mail.xidian.edu.cn}

\renewcommand{\shortauthors}{Trovato et al.}

\begin{abstract}

Large-model-driven embodied agents integrate foundation models with perception, reasoning, planning, and physical action, extending conventional model-level risks into embodied closed loops. Existing studies on their security and privacy remain fragmented across different system components and operational stages, making it difficult to understand how risks arise, propagate, and ultimately affect physical behavior or sensitive information. This survey presents a lifecycle-based analysis of security and privacy in large-model-driven embodied agents. We organize existing research into five stages: model construction and supply chain, multimodal input and interaction, semantic reasoning and task planning, action execution and physical feedback, and long-term deployment. Within this lifecycle, we systematically review representative attacks, defenses, and evaluation methods. Our analysis shows that attack entry, consequence realization, and defense intervention often occur at different stages of the embodied closed loop. It further reveals substantial gaps in end-to-end protection, real-world evaluation, and long-term privacy governance. This survey provides a unified perspective for understanding current progress and identifying critical directions for securing large-model-driven embodied agents.
\end{abstract}





\maketitle

\section{Introduction}

Large models are becoming central components of embodied agents and are reshaping the design of robotic systems. Earlier robotic systems typically decomposed perception, planning, and control into relatively fixed modules, with behavior determined by predefined procedures, task-specific models, and low-level controllers. In recent years, foundation models have increasingly been integrated into the perception, decision-making, and action processes of embodied systems, changing how they understand environments, make decisions, and execute tasks. Based on this shift, this survey uses the term \textit{large-model-driven embodied agents} to refer to agents in which large-model capabilities substantially participate in the embodied closed loop, covering typical robots and other embodied autonomous systems.

This change in system form also changes the boundary of security and privacy problems. Risks already present in ordinary LLMs may enter a longer system chain when embedded in an embodied closed loop. They can alter environmental perception, interfere with task planning, manipulate tool or API calls, and ultimately affect action trajectories and control commands. Since embodied systems involve perception, execution, and deployment, large-model risks coupled with these stages may extend from the information space into the physical world. The security of large-model-driven embodied agents therefore needs to be understood along the full system chain, rather than being confined to model inputs and outputs.

Privacy risks also expand with embodied deployment. Embodied agents continuously encounter user- and environment-related information during long-term operation, and such information often supports continual learning, decision optimization, and service provision. Privacy in large-model-driven embodied agents therefore cannot be limited to training-data leakage; it must also cover risks from long-term interaction, persistent sensing, and system deployment.

A systematic survey of security and privacy in large-model-driven embodied agents is both scientifically important and practically urgent. These agents are increasingly entering high-impact physical-world scenarios, where risks propagated through the embodied closed loop may exceed the model layer and affect the physical environment. Existing studies are distributed across different lifecycle stages, with differences in problem focus, attack entry points, defense locations, and evaluation settings. Organizing the literature only by model type, attack form, or application scenario makes it difficult to explain where problems arise, which components they affect, and what evidentiary boundaries support the corresponding claims.

Defense and evaluation studies face a similar difficulty. Some methods verify task safety before execution, some constrain trajectories or control commands at runtime, and others monitor risks during system operation. These methods are deployed at different parts of the embodied system and differ in target objects and evaluation goals. Consequently, existing research often characterizes local-stage security, but provides a limited view of how risks affect the complete embodied closed loop. Without a unified process-oriented perspective, it is difficult to compare defense coverage or relate different evaluation results.

Motivated by these observations, this survey adopts a lifecycle perspective to organize research on security and privacy in large-model-driven embodied agents. We first define the survey object and system boundary, restricting the scope to embodied-agent systems in which large-model capabilities substantially participate in the embodied closed loop. We then propose a five-stage lifecycle framework to organize existing studies. This framework characterizes system stages, attack and defense mechanisms, and evaluation methods from the perspective of system operation, thereby providing a clearer analytical structure for the field.

The main contributions of this survey are summarized as follows:
\begin{enumerate}
\item \textbf{We define the object boundary for security and privacy research on large-model-driven embodied agents.} We focus on embodied agents substantially driven by large-model capabilities, distinguish this topic from ordinary LLM security, software agent security, and traditional robotics safety, and extend the analysis to the system boundary that supports agent operation.

\item \textbf{We propose a five-stage lifecycle framework for large-model-driven embodied agents.} Starting from the closed-loop operation of embodied systems, we divide the lifecycle into five stages: model construction and supply chain, multimodal input and interaction, semantic reasoning and task planning, action execution and physical feedback, and deployment and human-centered impacts. This framework helps analyze where security and privacy risks arise, how they propagate, their attack consequences, and applicable defenses.

\item \textbf{We analyze attack and defense mechanisms and evaluation methods by lifecycle stage.} We categorize and discuss existing studies according to the lifecycle stage in which their main effect or impact occurs, and analyze risk formation, attack and defense mechanisms, and evaluation methods within different stages.

\item \textbf{We analyze cross-lifecycle risk propagation, defense composition, and open challenges.} We further examine cross-stage risk propagation and its implications for system governance, and discuss remaining challenges in defense composition, system-level evaluation, real-world validation, and long-term deployment governance.
\end{enumerate}

The remainder of this survey is organized as follows. Section~\ref{sec:background-and-scope} defines large-model-driven embodied agents and clarifies the boundaries among security, privacy, safety, robustness, and trustworthiness, while also comparing related surveys. Section~\ref{sec:taxonomy} introduces the five-stage lifecycle framework and three categories of research problems. Section~\ref{sec:lifecycle-security-analysis} provides the lifecycle-based security and privacy analysis. Sections~\ref{sec:model-construction-training-supply-chain-security}--\ref{sec:deployment-multi-agent-human-centered-impacts} provide stage-specific analyses, and Section~\ref{sec:cross-lifecycle-risk-propagation} discusses cross-lifecycle risk propagation. Section~\ref{sec:gaps-future-directions} identifies the remaining research gaps and outlines future directions. Section~\ref{sec:conclusion} concludes the survey.

\section{Background and Scope}
\label{sec:background-and-scope}

This section establishes the survey's object and analytical boundaries. We first positions the survey against related reviews, then distinguishes embodied agents, embodied agent systems, and large-model-driven embodied agents. We finally defines \textit{security}, \textit{privacy}, \textit{safety}, \textit{robustness}, and \textit{trustworthiness} as used throughout the survey.

\subsection{Related Work and Survey Positioning}

Existing surveys have provided important foundations for studying the security and privacy of large-model-driven embodied agents, while differing in their research scope and organizational granularity. Table~\ref{tab:related-work-survey-positioning} compares the closest studies by research scope, embodied-pipeline coverage, attack and defense classification, evaluation analysis, and security-privacy-deployment integration.

\begin{table}[t]
\caption{Comparison with closely related surveys}
\label{tab:related-work-survey-positioning}
\centering
\scriptsize
\setlength{\tabcolsep}{6.5pt}
\begin{tabularx}{\dimexpr\textwidth-6pt\relax}{lcccccc}
\toprule
\textbf{Survey} & \textbf{Broad LM-EA scope} & \textbf{Full pipeline} & \textbf{Attack} & \textbf{Defense} & \textbf{Evaluation} & \textbf{Sec-Priv-deployment} \tabularnewline
\midrule
Xing et al.~\cite{P003} & $\boldsymbol{\surd}$ & $\boldsymbol{\surd}$ & $\boldsymbol{\surd}$ & $\boldsymbol{\triangle}$ & $\boldsymbol{\surd}$ & $\boldsymbol{\triangle}$ \tabularnewline
\addlinespace[1.0ex]
Ma et al.~\cite{P1291} & $\boldsymbol{\triangle}$ & $\boldsymbol{\triangle}$ & $\boldsymbol{\times}$ & $\boldsymbol{\times}$ & $\boldsymbol{\times}$ & $\boldsymbol{\times}$ \tabularnewline
\addlinespace[1.0ex]
Huang et al.~\cite{P1305} & $\boldsymbol{\triangle}$ & $\boldsymbol{\times}$ & $\boldsymbol{\triangle}$ & $\boldsymbol{\triangle}$ & $\boldsymbol{\surd}$ & $\boldsymbol{\times}$ \tabularnewline
\addlinespace[1.0ex]
Li et al.~\cite{P1381} & $\boldsymbol{\triangle}$ & $\boldsymbol{\triangle}$ & $\boldsymbol{\triangle}$ & $\boldsymbol{\triangle}$ & $\boldsymbol{\surd}$ & $\boldsymbol{\times}$ \tabularnewline
\addlinespace[1.0ex]
Li et al.~\cite{P1732} & $\boldsymbol{\surd}$ & $\boldsymbol{\surd}$ & $\boldsymbol{\surd}$ & $\boldsymbol{\surd}$ & $\boldsymbol{\times}$ & $\boldsymbol{\triangle}$ \tabularnewline
\addlinespace[1.0ex]
Baraldi et al.~\cite{P1733} & $\boldsymbol{\triangle}$ & $\boldsymbol{\times}$ & $\boldsymbol{\times}$ & $\boldsymbol{\times}$ & $\boldsymbol{\surd}$ & $\boldsymbol{\times}$ \tabularnewline
\addlinespace[1.0ex]
Zeng et al.~\cite{P1811} & $\boldsymbol{\triangle}$ & $\boldsymbol{\times}$ & $\boldsymbol{\times}$ & $\boldsymbol{\times}$ & $\boldsymbol{\triangle}$ & $\boldsymbol{\times}$ \tabularnewline
\addlinespace[1.0ex]
Tan et al.~\cite{P5492} & $\boldsymbol{\surd}$ & $\boldsymbol{\surd}$ & $\boldsymbol{\times}$ & $\boldsymbol{\times}$ & $\boldsymbol{\surd}$ & $\boldsymbol{\surd}$ \tabularnewline
\addlinespace[1.0ex]
Wang et al.~\cite{P5496} & $\boldsymbol{\surd}$ & $\boldsymbol{\triangle}$ & $\boldsymbol{\times}$ & $\boldsymbol{\triangle}$ & $\boldsymbol{\times}$ & $\boldsymbol{\triangle}$ \tabularnewline
\addlinespace[1.0ex]
\textbf{This survey} & $\boldsymbol{\surd}$ & $\boldsymbol{\surd}$ & $\boldsymbol{\surd}$ & $\boldsymbol{\surd}$ & $\boldsymbol{\surd}$ & $\boldsymbol{\surd}$ \tabularnewline
\bottomrule
\end{tabularx}

\smallskip
\parbox{\textwidth}{\footnotesize
\textit{{Note:} \textbf{Broad LM-EA scope}: broad coverage of large-model-driven embodied agents; \textbf{Full pipeline}: full embodied pipeline coverage; \textbf{Attack}: structured attack taxonomy across the embodied lifecycle; \textbf{Defense}: structured defense taxonomy across the embodied lifecycle; \textbf{Evaluation}: evaluation mechanisms and benchmarks; \textbf{Sec-Priv-deployment}: integrated treatment of security, privacy, and deployment. ($\boldsymbol{\surd}$) indicates systematic coverage or use as a major organizing dimension; ($\boldsymbol{\triangle}$) indicates partial coverage without a systematic taxonomy; ($\boldsymbol{\times}$) indicates absent, marginal, or non-systematic coverage.}}
\end{table}

Xing et al.~\cite{P003} discuss vulnerabilities, attack surfaces, evaluation, and benchmarks in embodied AI, but provide limited organization of defenses and lifecycle stages. Ma et al.~\cite{P1291} frame failures as mismatches among LLMs, CPS, and physical environments. This system view explains composite failures, yet does not develop a closed-loop attack and defense taxonomy. Huang et al.~\cite{P1305} review threats, defenses, and evaluation for LLM-controlled robotics, but emphasize LLM integration and planning over perception, interaction, physical execution, and feedback. Li et al.~\cite{P1381} synthesize attacks, defenses, and evaluation for VLA safety, leaving broader large-model-driven embodied systems outside their scope.

Other surveys address embodied-AI safety, world-model safety, and trustworthy embodied AI. Li et al.~\cite{P1732} organize attacks and defenses along the embodied operating loop but treat evaluation less extensively. This organization captures operational interdependence without comparing the evidentiary reach of different evaluation settings. Baraldi et al.~\cite{P1733} and Zeng et al.~\cite{P1811} concentrate on future-state prediction, simulation, and world-model-specific safety pathologies rather than broader embodied attack and defense problems. Tan et al.~\cite{P5492} propose a maturity model and system-level principles, benchmarks, and metrics without a fine-grained attack and defense taxonomy. Wang et al.~\cite{P5496} examine security, privacy, and reliability across embodied intelligence, IoT, and cloud-edge architectures, treating them mainly as system-enabling dimensions rather than jointly classifying attacks, defenses, and evaluation.

Prior surveys lack a unified security-and-privacy account that organizes attacks, defenses, and evaluation across the full lifecycle of large-model-driven embodied agents. The analytical gap matters because studies with different units of analysis cannot be consistently compared by risk entry, defensive intervention, or evidentiary reach. We address it by defining the object explicitly and using five stages to locate each study's intervention, affected components, and evidence boundary.

\subsection{Large-Model-Driven Embodied Agents}

Embodied agents interpret environments and task goals and affect the external world through action. They operate through a closed loop of perception, reasoning, planning, action, and feedback. Unlike traditional control systems, they combine sensing and feedback with task-level environmental understanding and decision planning. Their risks therefore arise not only within models, but also across information and control flows. An embodied agent system comprises the agent and its supporting software and hardware, including sensors, foundation models, communication mechanisms, and infrastructure. This broader boundary captures dependencies through which model-level problems can propagate and become physical consequences.

\begin{figure}[t]
  \centering
  \includegraphics[width=0.96\textwidth]{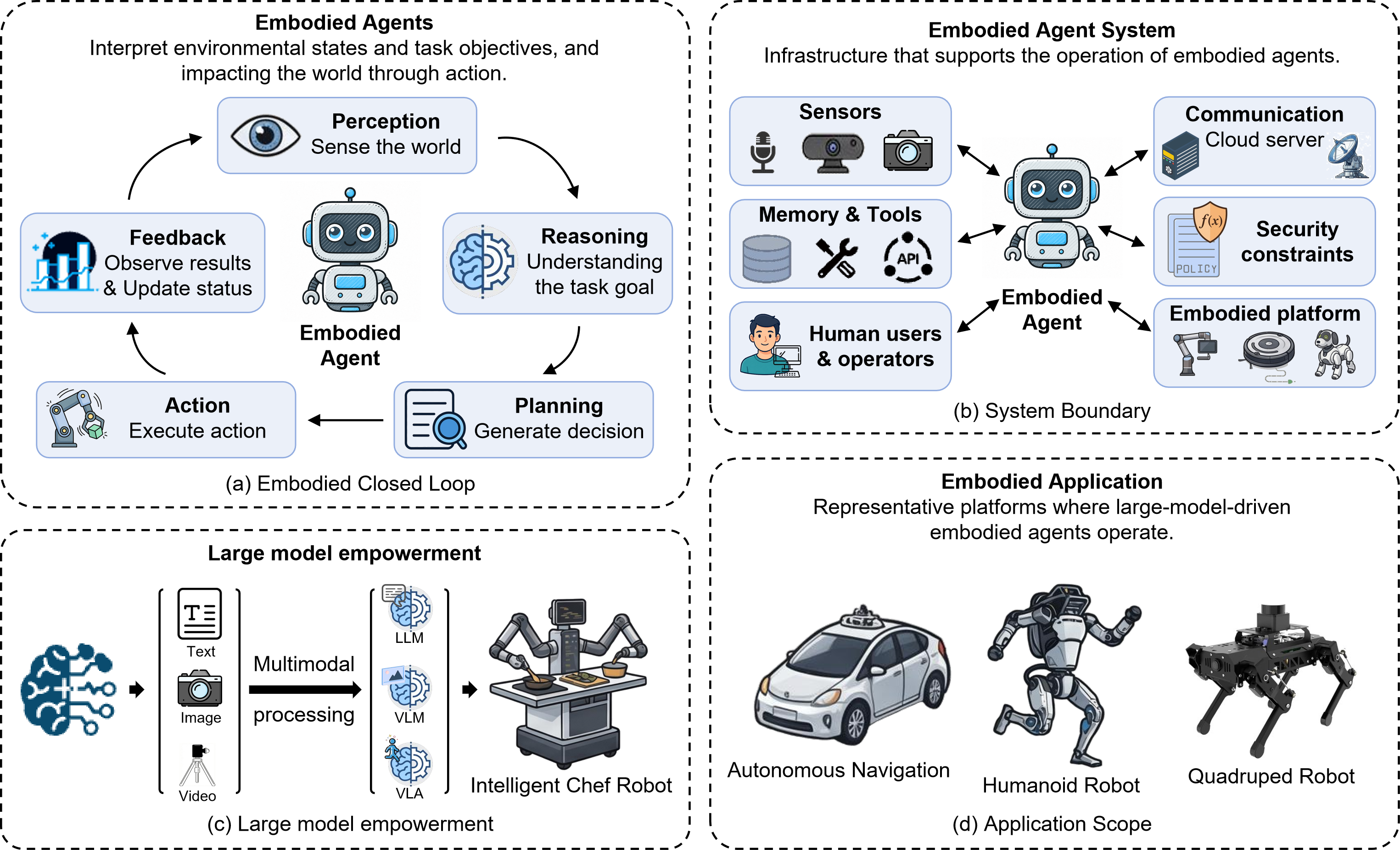}
  \caption{Conceptual scope and system boundary of large-model-driven embodied agents.}
  \Description{A conceptual illustration of the scope and system boundary of large-model-driven embodied agents, showing embodied agents operating in a closed loop of perception, reasoning, planning, action, and feedback, together with supporting components such as sensors, foundation models, communication mechanisms, and other infrastructure.}
  \label{fig:LMEA-system-boundary}
\end{figure}

This survey focuses on embodied agents whose core capabilities are substantially driven by large models. We refer to them as large-model-driven embodied agents and use embodied agent systems as the system boundary for security and privacy analysis. In these agents, environmental understanding, task reasoning, decision planning, and action generation are often handled by one or more large models rather than traditional rule systems or task-specific control modules. System behavior therefore depends heavily on the model's knowledge representation, generalization ability, and contextual reasoning. As language, multimodal, and robotic foundation models are increasingly integrated into embodied closed loops, large models are shifting from auxiliary components to central engines of behavior generation and task execution. Consequently, information flows, control flows, and authority boundaries are increasingly organized around large models, making model capability, training data, and inference processes important determinants of embodied behavior. Prior work shows that large models have entered robotic perception, planning, and decision-making, with roles expanding from language interfaces to multimodal understanding and action generation~\cite{P1552,P5493,P1695}. Figure~\ref{fig:LMEA-system-boundary} illustrates the conceptual scope and system boundary of large-model-driven embodied agents.

Under this definition, the survey covers embodied agents and embodied agent systems driven by large-model capabilities. For domains such as autonomous driving, UAVs, and multi-robot systems, the focus is on the large-model-driven embodied agents within these systems and on security and privacy problems arising in their operational loops.

\subsection{Security, Privacy, Safety, Robustness, and Trustworthiness}

This survey centers on \textit{security} and \textit{privacy}, while recognizing their close relationship with \textit{safety}, \textit{robustness}, and \textit{trustworthiness} in \textbf{large-model-driven embodied agents}.

\textit{Security} refers to a system's ability to preserve intended functions, authority boundaries, and execution behavior under malicious interference or unauthorized influence. Ordinary LLM security mainly examines attacker manipulation or exploitation of model behavior~\cite{P122}, whereas traditional robot and CPS security emphasizes operational control and infrastructure protection~\cite{P5533}. Once large models are integrated into embodied closed loops, these two types issues of security become intertwined. Attackers may manipulate model inputs and reasoning, but may also affect communication, planning, and execution chains.

\textit{Privacy} refers to a system's ability to protect and govern sensitive information and personal data, and to prevent improper access or leakage. LLM privacy research mainly studies leakage during model training, user interaction, and model use~\cite{P145}. In embodied agents, the privacy boundary expands because systems continuously encounter real-world environmental data and user interaction records~\cite{P5547}. Privacy in this survey therefore includes not only training-data and user-information leakage, but also protection during continuous sensing, memory storage, and deployed operation.

\textit{Safety} concerns whether system behavior may cause unacceptable real-world consequences. \textit{Robustness} concerns whether the system can maintain stable and reliable behavior under complex conditions. \textit{Trustworthiness} is a broader concept describing whether the system is worthy of user and societal trust as a whole~\cite{P5485}. The following sections focus on security and privacy. Safety, robustness, and trustworthiness are used mainly to characterize the behavioral consequences induced by security and privacy problems, as well as the requirements they create for deployment and governance.

\section{Five-Stage Lifecycle Taxonomy}
\label{sec:taxonomy}

We organize security and privacy research on large-model-driven embodied agents from a lifecycle perspective. Compared with classifications based on model type, attack form, or system component, a lifecycle view better captures how embodied agents operate from model construction to long-term deployment. Existing studies often focus on a specific model architecture, attack mechanism, or system module; while useful for local analysis, these views make it difficult to form a unified picture. Large-model-driven embodied agents combine multiple models and system components and interact with the environment through a continuous perception-decision-execution loop. Their security and privacy problems therefore span model construction, input processing, task planning, action execution, and long-term deployment. Since attack surfaces, defense mechanisms, and evaluation priorities differ by stage, a process-oriented framework is needed to clarify which stage a study addresses, what it affects, and what evidence supports its conclusions.

Existing research also shows clear stage-specific patterns. AttackVLA mainly studies attack evaluation in model construction and training~\cite{P1436}. CHAI treats visual text in the physical environment as an entry point for multimodal input attacks~\cite{P5541}. SafeGate and language-conditioned safety filtering propose defense mechanisms for pre-execution verification and runtime control filtering, respectively~\cite{P1280,P1311}. Work on MITM attacks against an LLM-enabled vacuum robot reveals system-level risks in deployment-stage communication links~\cite{P5505}. Although these studies target different technical objects and problems, they can be mapped to specific parts of the embodied-agent lifecycle. This mapping motivates the lifecycle-based organization used in this survey.

Based on this observation, we adopt a five-stage lifecycle framework. The framework follows the system's operational process and organizes studies by the stage at which their main security or privacy effect occurs. By highlighting the core security problem at each stage, it supports clearer comparison of attacks, defenses, and evaluations.

\begin{figure}[t]
  \centering
  \includegraphics[width=0.93\textwidth]{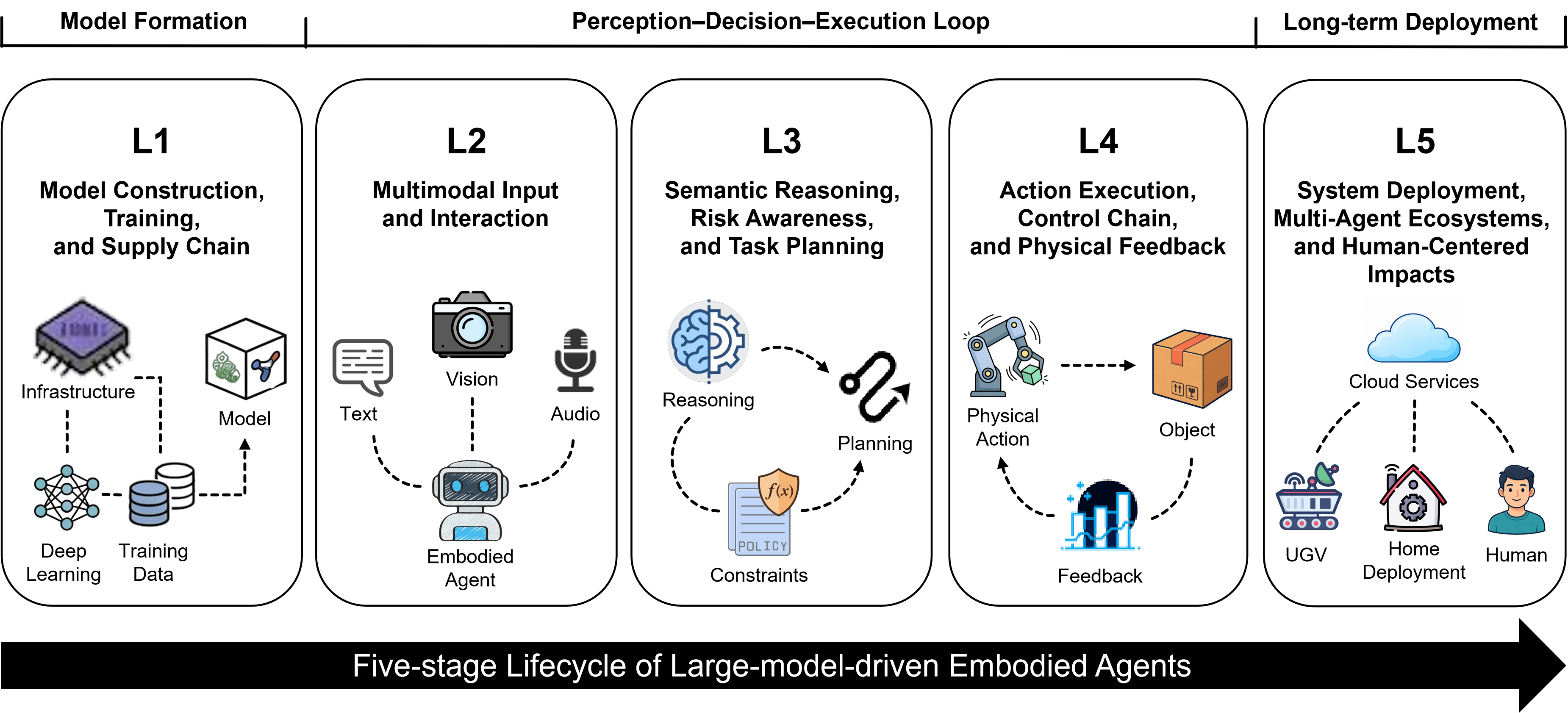}
  \caption{Five-stage lifecycle framework for large-model-driven embodied agents. By incorporating model construction (L1) and long-term deployment ecosystems and impacts (L5), the framework extends the conventional perception--decision--execution loop (L2--L4) into a complete embodied-agent lifecycle.}
  \Description{The figure illustrates a five-stage lifecycle of large-model-driven embodied agents. From left to right, the stages cover model construction, training, and supply-chain resources; multimodal input and interaction through language, vision, voice, and robotic interfaces; semantic reasoning, risk awareness, and task planning with policies and safety constraints; action execution and physical feedback through robotic manipulation and performance monitoring; and system deployment, multi-agent ecosystems, and human-centered impacts involving cloud services, users, homes, and external environments.}
  \label{fig:five-stage-lifecycle}
\end{figure}

\subsection{Lifecycle Stages}

We divide the lifecycle of large-model-driven embodied agents into five stages: L1 Model Construction, Training, and Supply Chain; L2 Multimodal Input and Interaction; L3 Semantic Reasoning, Risk Awareness, and Task Planning; L4 Action Execution, Control Chain, and Physical Feedback; and L5 System Deployment, Multi-Agent Ecosystems, and Human-Centered Impacts, as illustrated in Figure~\ref{fig:five-stage-lifecycle}. Table~\ref{tab:five-stage-lifecycle-taxonomy} summarizes the main model roles and representative problems in each stage.

\begin{table}[t]
\centering
\caption{Five-stage lifecycle taxonomy of large-model-driven embodied agents}
\label{tab:five-stage-lifecycle-taxonomy}
\scriptsize
\setlength{\tabcolsep}{3pt}
\begin{tabularx}{\dimexpr\textwidth-6pt\relax}{
L{0.05\textwidth}
L{0.22\textwidth}
L{0.39\textwidth}
Y
}
\toprule
\textbf{Stage} & \textbf{Lifecycle stage} & \textbf{Major model families} & \textbf{Representative security and privacy concerns} \tabularnewline
\midrule

L1 &
Model Construction, Training, and Supply Chain &
VLA and robot foundation models learn action generation from multimodal data, demonstrations, action representations, and model updates. &
Backdoors, supply-chain compromise, training-data inference, capability removal and unlearning gaps. \tabularnewline
\addlinespace[1.0ex]

L2 &
Multimodal Input and Interaction &
LLMs process language interaction; VLMs/MLLMs interpret scenes and environmental text; VLA models pass multimodal inputs into action generation. &
Sensor attacks, adversarial perception, environmental prompt injection, jailbreak. \tabularnewline
\addlinespace[1.0ex]

L3 &
Semantic Reasoning, Risk Awareness, and Task Planning &
LLMs act as planners and tool selectors; VLMs/MLLMs provide grounded scene information; world models support future-state prediction. &
Unsafe planning, policy-executable jailbreak, weak risk awareness, unsafe tool/API use, incomplete safety-constraint translation. \tabularnewline
\addlinespace[1.0ex]

L4 &
Action Execution, Control Chain, and Physical Feedback &
VLA and robot foundation models generate action sequences, trajectories, or policy outputs for physical execution. &
Action manipulation, unsafe trajectories, runtime constraint violation, recovery failure. \tabularnewline
\addlinespace[1.0ex]

L5 &
System Deployment, Multi-Agent Ecosystems, and Human-Centered Impacts &
LLMs and cloud models support deployed decision-making, cloud interaction, long-term operation, and system coordination. &
Communication hijacking, unsafe propagation, access-control failure, long-term privacy leakage. \tabularnewline

\bottomrule
\end{tabularx}
\end{table}

L1 concerns security and privacy before a model enters the embodied closed loop and during later updates. For VLA and robot foundation models, training data, action representations, and demonstration trajectories directly shape how robot actions are generated from multimodal context, making them a key security boundary. Existing studies show that poisoned demonstration trajectories can bind triggers to specific action behaviors and induce robots to execute attacker-specified action sequences at inference time~\cite{P1436}.

L2 is the stage where embodied agents receive external information from users and physical environments. VLMs and MLLMs handle scene understanding and environmental text interpretation, LLMs process language interaction, and VLA models may pass inputs into the action-generation chain. CHAI shows that misleading text in the physical environment can alter an LVLM's scene interpretation and affect downstream decisions~\cite{P5541}.

L3 covers the understanding of environmental states and task goals, as well as reasoning and plan generation. LLMs often serve as planners or tool selectors, VLMs and MLLMs provide grounded scene information, and world models may support future-state prediction. The central question is whether a semantically plausible output can become a plan or policy with harmful embodied consequences. POEX shows that robotic attack targets can shift from harmful text to harmful yet executable policies~\cite{P5498}.

L4 covers the conversion of high-level decisions into actions and control commands. VLA and robot foundation models often generate action sequences, trajectories, or policy outputs. Because L4 directly connects to the physical environment, planning errors, action manipulation, or abnormal control can become concrete robot behavior. Language-conditioned safety filtering maps language constraints into structured safety specifications and constrains robot behavior through a runtime control filter~\cite{P1311}. L4 is therefore both where physical consequences materialize and where runtime defenses can intervene.

L5 concerns system-level security and privacy after embodied agents enter long-term operating environments. Its scope includes continuous operation and long-term interaction after deployment. Research on an LLM-enabled vacuum robot shows that an attacker can modify control instructions and user feedback by tampering with communication between the robot and the cloud model~\cite{P5505}.

Together, the five lifecycle stages cover the process from model formation to practical deployment and correspond to distinct security and privacy challenges. This framework places studies from different parts of the system lifecycle into a unified view and supports comparison of attack surfaces, defense mechanisms, and evaluation methods by stage.

\subsection{Research Problem Types and Lifecycle-by-problem Matrix}

On top of the lifecycle framework, we classify existing work into three research problem types according to its primary contribution: \textit{Attacks}, \textit{Defenses}, and \textit{Evaluation}. These categories correspond to attack mechanisms and risk exposure, defense and mitigation methods, and evaluation benchmarks or experimental validation.

\textit{Attacks} concern how security and privacy risks arise in large-model-driven embodied agents, how attacks proceed, and what effects they produce. This line of work studies both the exploitation of system weaknesses and the downstream consequences of such interference. For security, major topics include training-stage attacks, input manipulation, and interference with planning or execution chains. For privacy, the focus is on data leakage, sensitive information exposure, and privacy inference. The central goal is to reveal threats, attack surfaces, and potential consequences in embodied agents.

\textit{Defenses} concern mechanisms for detecting, mitigating, blocking, or governing these risks. Existing studies not only identify and filter attacks, but also introduce safety constraints during system operation to limit risk propagation. Examples include task verification, risk assessment, runtime supervision for generated plans and actions, and access management or privacy governance for long-term deployment~\cite{P1280,P1311}. The core question is how to improve safety, robustness, and privacy protection while preserving task capability and interaction efficiency.

\textit{Evaluation} concerns evaluation methods, benchmarks, and assessment frameworks for security and privacy problems. Because embodied agents involve a full chain from perception to execution, conventional large-model evaluation alone cannot capture system-level risk. Recent work has therefore begun to develop embodied benchmarks and experimental frameworks for assessing attacks, defenses, and overall trustworthiness in embodied settings~\cite{P1286,P5485}.

Combining the five-stage lifecycle framework with the three research problem types yields a lifecycle-by-problem matrix. Table~\ref{tab:lifecycle-problem-matrix} shows how existing studies are distributed across lifecycle stages and problem types.

\begin{table}[t]
\centering
\caption{Lifecycle-by-problem taxonomy matrix}
\label{tab:lifecycle-problem-matrix}
\tiny
\setlength{\tabcolsep}{3pt}
\begin{tabularx}{\dimexpr\textwidth-6pt\relax}{
L{0.21\textwidth}
L{0.25\textwidth}
Y
L{0.20\textwidth}
}
\toprule
\textbf{Lifecycle stage} & \textbf{Attacks} & \textbf{Defenses} & \textbf{Evaluation} \tabularnewline
\midrule
L1 Model Construction, Training, and Supply Chain &
\cite{P1322}\textsuperscript{\textdagger\textsection}; \cite{P1422}\textsuperscript{\textdagger}; \cite{P1424}\textsuperscript{\textdagger}; \cite{P1436}\textsuperscript{\textdagger}; \cite{P1441}\textsuperscript{\textdagger}; \cite{P1750}; \cite{P5484}; \cite{P5503}; \cite{P5510}\textsuperscript{\textdagger}; \cite{P5526}; \cite{P5539}\textsuperscript{\textdagger} &
\cite{P1402}; \cite{P1457}; \cite{P2232}\textsuperscript{\textdaggerdbl}; \cite{P5512} &
-- \tabularnewline
\thinmidrule

L2 Multimodal Input and Interaction &
\cite{P1385}; \cite{P1392}; \cite{P1433}\textsuperscript{\textdagger}; \cite{P1438}; \cite{P1442}; \cite{P1458}; \cite{P1743}; \cite{P2478}; \cite{P4822}; \cite{P5483}\textsuperscript{\textdagger}; \cite{P5490}; \cite{P5507}\textsuperscript{\textdagger}; \cite{P5508}; \cite{P5509}; \cite{P5527}; \cite{P5541}\textsuperscript{\textdagger} &
\cite{P1306}; \cite{P1403}; \cite{P1415}; \cite{P1826}; \cite{P219}; \cite{P4845}\textsuperscript{\textsection}; \cite{P5073} &
\cite{P1769}\textsuperscript{\textdaggerdbl}; \cite{P3452}\textsuperscript{\textsection}; \cite{P5486}; \cite{P5511}; \cite{P5523}; \cite{P5525}\textsuperscript{\textdagger}; \cite{P5530}; \cite{P5504} \tabularnewline
\thinmidrule

L3 Semantic Reasoning, Risk Awareness, and Task Planning &
\cite{P5489}; \cite{P5498}\textsuperscript{\textdagger}; \cite{P5506}\textsuperscript{\textdagger} &
\cite{P1274}; \cite{P1280}; \cite{P1316}; \cite{P1395}; \cite{P1421}; \cite{P1443}; \cite{P1454}; \cite{P1561}; \cite{P1569}; \cite{P1834}\textsuperscript{\textdaggerdbl}; \cite{P1862}\textsuperscript{\textdaggerdbl}; \cite{P2468}\textsuperscript{\textdaggerdbl}; \cite{P2500}; \cite{P2566}; \cite{P3632}\textsuperscript{\textdaggerdbl}; \cite{P3874}; \cite{P5168}\textsuperscript{\textdaggerdbl}; \cite{P5242}; \cite{P5482}; \cite{P5487}; \cite{P5501}; \cite{P5513}; \cite{P5524}\textsuperscript{\textdagger}; \cite{P5540}; \cite{P5542} &
\cite{P1281}; \cite{P1286}\textsuperscript{\textdagger}; \cite{P1383}; \cite{P4190}\textsuperscript{\textdaggerdbl}; \cite{P4781}; \cite{P5133}\textsuperscript{\textdaggerdbl}; \cite{P5488}; \cite{P5499}; \cite{P5500}; \cite{P5502}; \cite{P5522}\textsuperscript{\textsection} \tabularnewline
\thinmidrule

L4 Action Execution, Control Chain, and Physical Feedback &
\cite{P116} &
\cite{P1299}; \cite{P1311}\textsuperscript{\textdagger}; \cite{P1397}; \cite{P1426}; \cite{P1434}; \cite{P1562}\textsuperscript{\textdaggerdbl}; \cite{P1689}\textsuperscript{\textdagger}; \cite{P1690}; \cite{P1736}\textsuperscript{\textdaggerdbl}; \cite{P182}; \cite{P2697}\textsuperscript{\textdagger\textdaggerdbl}; \cite{P4996}\textsuperscript{\textdaggerdbl}; \cite{P5034}\textsuperscript{\textdaggerdbl}; \cite{P5461}\textsuperscript{\textdaggerdbl}; \cite{P5543} &
\cite{P1382}\textsuperscript{\textdagger}; \cite{P1410} \tabularnewline
\thinmidrule

L5 System Deployment, Multi-Agent Ecosystems, and Human-Centered Impacts &
\cite{P1275}\textsuperscript{\textdagger}; \cite{P1279}\textsuperscript{\textdagger\textdaggerdbl}; \cite{P3593}\textsuperscript{\textdaggerdbl}; \cite{P5505}\textsuperscript{\textdagger} &
\cite{P008}; \cite{P1324}\textsuperscript{\textdagger}; \cite{P2230}\textsuperscript{\textdagger\textdaggerdbl}; \cite{P2497}\textsuperscript{\textdagger}; \cite{P4343}\textsuperscript{\textdagger\textdaggerdbl}; \cite{P4762} &
\cite{P5485}\textsuperscript{\textdagger}; \cite{P148}; \cite{P1568}\textsuperscript{\textsection} \tabularnewline
\bottomrule
\end{tabularx}

\smallskip
\parbox{\dimexpr\textwidth-6pt\relax}{\footnotesize
\textit{\textbf{Note:} Studies are assigned by their primary contribution: attack construction or validation, defense design, or evaluation resource. (\textdagger) indicates explicit cross-lifecycle propagation evidence; (\textsection) indicates model-output or offline-only validation; (\textdaggerdbl) indicates trustworthiness, privacy, functional-safety, or conceptual boundary work.}}
\end{table}

The matrix shows that input attacks and runtime defenses have formed relatively clear research lines, whereas training and supply-chain security and long-term deployment governance remain less developed.

\subsection{Cross-lifecycle Issues}

The five-stage lifecycle framework identifies the primary stage at which a security or privacy problem acts within an embodied-agent lifecycle. However, not every problem can be fully explained within a single stage. Some attacks enter at one stage and are triggered or amplified downstream, while some defenses are deployed after the attack entry point to block later risk propagation.

Existing studies show that the attack entry point often differs from the stage where consequences appear. Training-stage data poisoning or backdoor insertion may only manifest as abnormal behavior during later action generation~\cite{P1436}. Physical prompts in the environment may first affect scene understanding and then interfere with task planning and action execution~\cite{P5541}. Message tampering in deployment-stage communication can bypass internal model mechanisms and directly modify control instructions or user feedback~\cite{P5505}. Correspondingly, many defenses intervene not at the attack entry point, but at the planning or execution layers through formal verification, runtime filtering, and related mechanisms to block or mitigate upstream risks~\cite{P5524,P1311}.

For problems with clear cross-stage propagation, single-stage analysis cannot fully capture the attack path, affected scope, or defense composition. Chapter~9 therefore revisits these problems from a cross-lifecycle perspective and analyzes their risk propagation paths and implications for system governance.

\section{Lifecycle Security and Privacy Analysis}
\label{sec:lifecycle-security-analysis}

\subsection{Model Construction, Training, and Supply Chain Security}
\label{sec:model-construction-training-supply-chain-security}

Model construction, training, and supply chain form the foundational stage before large-model-driven embodied agents enter embodied closed loops. This stage also includes model updates and external component integration. Training data and model design jointly determine which behaviors the system can learn and how they may be activated in later tasks. Risks introduced here may remain latent for a long period and later appear as action deviation, task failure, or privacy leakage under specific conditions.

This stage covers model training, updating, and component integration. Its security boundary includes not only training data and model parameters, but also action representations, training infrastructure, and external model or service interfaces. For VLA and robot foundation models, these elements directly affect how a model generates agent actions from multimodal context and should be treated as critical security assets. Training-data poisoning does not only bias textual responses; it may also create abnormal associations between triggers and action sequences, trajectory structures, or control-related representations. Modular embodied systems and cloud-edge deployment further expand the supply-chain boundary. Distributed training and federated fine-tuning also introduce risks through parameter transmission and node coordination. L1 security should therefore cover training services and continuous-update infrastructure, rather than only the main policy model.

\subsubsection{Attacks}

L1 attacks occur during training, fine-tuning, model distribution, or component integration. They aim to implant persistent malicious behaviors before deployment or infer sensitive information from training artifacts. In embodied systems, these attacks usually target action generation, trajectory control, or inter-module information flow, rather than merely textual outputs. Existing attacks mainly fall into three directions: backdoor injection, supply-chain compromise, and training-data inference. Table~\ref{tab:l1-attack-mechanisms} summarizes representative mechanisms in this stage.

\begin{table}[t]
\centering
\caption{Attack mechanisms in model construction, training, and supply-chain security}
\label{tab:l1-attack-mechanisms}
\scriptsize
\setlength{\tabcolsep}{3pt}
\begin{tabularx}{\dimexpr\textwidth-6pt\relax}{
L{0.17\textwidth}
L{0.19\textwidth}
Y
L{0.22\textwidth}
}
\toprule
\textbf{Attack Type} &
\textbf{Attack Methods} &
\textbf{Key Characteristics / Impact} &
\textbf{Datasets / Platforms} \tabularnewline
\midrule

Backdoor attacks &
BadVLA~\cite{P5526} &
Decouples trigger learning from clean-task optimization, enabling latent backdoors while largely preserving normal task behavior. &
LIBERO-Spatial; LIBERO-Object; LIBERO-Goal; LIBERO-Long; SimplerEnv. \tabularnewline
\addlinespace[1.0ex]

&
BackdoorVLA~\cite{P1436} &
Binds visual/textual triggers to attacker-specified long-horizon action trajectories, shifting the target from task failure to precise action control. &
LIBERO-Object; LIBERO-Spatial; LIBERO-Goal; LIBERO-10; 7-DoF Franka Emika arm. \tabularnewline
\addlinespace[1.0ex]

&
State Backdoor~\cite{P1422} &
Uses the robot initial state as a trigger, exposing proprioceptive state as a stealthy backdoor surface beyond visual prompts. &
LIBERO; SO101 robotic arm. \tabularnewline
\addlinespace[1.0ex]

&
DropVLA~\cite{P1441} &
Targets reusable low-level action primitives; open-gripper hijacking leads to object drop or grasp failure. &
LIBERO-Spatial; LIBERO-Goal; 7-DoF Franka Emika arm. \tabularnewline
\addlinespace[1.0ex]

&
SILENTDRIFT~\cite{P1424} &
Exploits action chunking and delta-pose integration, causing small smooth deviations to accumulate into stealthy trajectory failure. &
LIBERO. \tabularnewline
\addlinespace[1.0ex]

&
GoBA~\cite{P5510} &
Uses physical objects as triggers and links them to goal-oriented backdoor actions, making the attack visually grounded and task-directed. &
BadLIBERO; LIBERO. \tabularnewline
\addlinespace[1.0ex]

&
BEAT~\cite{P5539} &
Learns visual-object triggers for VLM-based embodied agents, enabling trigger-conditioned multi-step policy switching. &
VAB-OmniGibson; EB-ALFRED. \tabularnewline
\midrule

Supply-chain attacks &
TrojanRobot~\cite{P5503} &
Injects a malicious VLM/LVLM module into the LLM-to-VLM pathway, corrupting grounding without modifying the main policy. &
UR3e; myCobot 280-Pi. \tabularnewline
\midrule

Training-data inference attacks &
VLA membership inference attacks~\cite{P1322} &
Infers sample or trajectory membership from action errors and temporal dynamics, revealing privacy leakage in embodied demonstrations. &
LIBERO. \tabularnewline

\bottomrule
\end{tabularx}
\end{table}

\textbf{Backdoor attacks.} Backdoor attacks establish hidden associations between trigger conditions and target behaviors during model construction. The model preserves normal performance on clean inputs, but switches to attacker-specified behaviors when the trigger appears. For VLA and robot foundation models, the payload is often embedded in the action policy, making its consequence closer to physical execution than to an incorrect textual response.

BadVLA~\cite{P5526} and BackdoorVLA~\cite{P1436} illustrate representative VLA backdoor mechanisms. BadVLA targets Training-as-a-Service settings and uses Objective-Decoupled Optimization to separate trigger features from clean features while maintaining normal task performance. BackdoorVLA binds visual or textual triggers to attacker-specified long-horizon action trajectories, causing the model to generate targeted action sequences at inference time.

Embodied backdoors can also exploit state and action structures in robotic policies. State Backdoor~\cite{P1422} uses the robot initial state as a trigger under a specific proprioceptive state. DropVLA~\cite{P1441} compresses the payload into low-level action primitives and manipulates short-window action labels to induce open-gripper behavior, causing grasp failure or object dropping. SILENTDRIFT~\cite{P1424} exploits action chunking and delta-pose representation to implant smooth trajectory drift at critical execution stages, where small deviations accumulate into task failure. State representations, action windows, and trajectory structures therefore become backdoor trigger surfaces beyond explicit image patches or textual phrases.

Visual scenes can also serve as trigger surfaces. GoBA~\cite{P5510} uses physical objects in real-world as triggers and associates them with malicious goals through demonstration trajectories. BEAT~\cite{P5539} targets VLM-based embodied agents by using supervised fine-tuning and Contrastive Trigger Learning to associate visual triggers with multi-step malicious policies. These attacks rely on visible objects and scene semantics, allowing L1 backdoors to be activated by later input conditions.

\textbf{Supply-chain attacks.} Supply-chain attacks target external modules and service dependencies in model construction. Rather than directly poisoning the main policy model, an attacker may compromise a perception model, an LVLM interface, or an MLaaS service, and then affect robot behavior through component interaction. In TrojanRobot~\cite{P5503}, an attacker without access to the target policy model's training data or weights can insert a malicious VLM/LVLM module into the robot policy pipeline, changing visual grounding and affecting target localization and action execution. External model components and service interfaces therefore become part of the robot behavior control chain.

\textbf{Training-data inference attacks.} Training-data inference attacks do not directly manipulate robot behavior, but infer whether a sample or demonstration was used for training by observing the trained model's outputs. Membership inference attacks against VLAs distinguish sample-level and trajectory-level inference~\cite{P1322}, using action errors, trajectory smoothness, curvature, and related signals. Robot demonstrations may leave observable traces in action outputs. In household, assistive, and long-term interaction scenarios, such trajectories may contain private schedules, object layouts, and user behavior patterns, creating clear privacy risks.

The common risk of L1 attacks is that abnormal behavior bindings formed during model construction can persist into later lifecycle stages. State representations, action labels, and external modules may all carry these bindings and affect action generation or behavior decisions once runtime conditions activate them.

\subsubsection{Defenses}

L1 defenses aim to reduce latent risks before a model enters the embodied closed loop. They also seek to improve controllability during model updating and component integration. Existing work mainly explores safety alignment, post-training behavior governance, and training-infrastructure protection. These methods mitigate part of the risk, but they do not yet systematically detect VLA-specific backdoors or supply-chain compromise. Table~\ref{tab:l1-defense-mechanisms} summarizes the representative defense mechanisms in this stage.

\begin{table}[t]
\centering
\caption{Defense mechanisms in model construction, training, and supply-chain security}
\label{tab:l1-defense-mechanisms}
\scriptsize
\setlength{\tabcolsep}{3pt}
\begin{tabularx}{\dimexpr\textwidth-6pt\relax}{
L{0.26\textwidth}
L{0.18\textwidth}
L{0.22\textwidth}
Y
}
\toprule
\textbf{Defense Type} &
\textbf{Defense Methods} &
\textbf{Verification Scenario} &
\textbf{Datasets / Platforms} \tabularnewline
\midrule

Safety alignment and behavior correction &
SafeVLA~\cite{P1457} &
Simulation environment &
Safety-CHORES; AI2-THOR; iTHOR; ProcTHOR; RoboTHOR; dual Realman RM75-6F arms. \tabularnewline
\addlinespace[1.0ex]

&
TakeVLA~\cite{P1402} &
Simulation environment &
Bench2Drive; CARLA. \tabularnewline
\midrule

Capability removal and unlearning &
VLA-Forget~\cite{P5512} &
Non-embodied evaluation &
Open X-Embodiment; lerobot/pusht\_image (PushT). \tabularnewline
\midrule

Training-infrastructure protection &
DE-SPFF~\cite{P2232} &
Non-embodied evaluation &
Semantic Drone Dataset. \tabularnewline

\bottomrule
\end{tabularx}
\end{table}

\textbf{Safety alignment and behavior correction.} Safety alignment methods aim to reduce dangerous behavior during training or post-training. SafeVLA~\cite{P1457} formulates VLA safety alignment as a constrained Markov decision process and uses Safety-CHORES with safety predicates to incorporate safety requirements into policy optimization. TakeVLA~\cite{P1402} uses expert takeover data to post-train a driving VLA, enabling the model to learn pre-accident risk signals and improve closed-loop safety decisions. These methods mainly reduce the model's tendency toward unsafe behavior, but still lack direct detection capability for deliberate training poisoning and malicious supply-chain components.

\textbf{Capability removal and unlearning.} Capability governance concerns removing undesired behaviors or sensitive knowledge after training. VLA-Forget~\cite{P5512} formulates embodied-model forgetting as a balance among target behavior removal, perception preservation, and reasoning preservation, and performs capability removal through component-level selective updates. This direction is useful for lifecycle behavior governance, especially when a model has acquired dangerous skills or retained sensitive information. However, current evidence remains largely empirical and cannot strictly prove that a demonstration trajectory, object association, or dangerous behavior has been completely removed.

\textbf{Training-infrastructure protection.} Training-infrastructure protection addresses privacy and integrity risks in distributed training and model updating. SPFF and DE-SPFF~\cite{P2232} use encryption and proxy re-encryption to protect parameter transmission during federated LLM fine-tuning in UAV swarm networks, while improving robustness under node dropout. These methods reduce leakage and tampering risks during model updating, but mainly protect training communication and parameter aggregation. They do not replace data provenance, backdoor auditing, or component-level supply-chain certification.

\subsubsection{Evaluation}

Backdoor studies usually report attack success rate, clean-task success rate, and trigger stealthiness. Privacy-inference studies ask whether action outputs leak membership of training samples or demonstration trajectories. Defense studies evaluate safety alignment, capability removal, and parameter protection through task performance, safety violations, or recovery capability. Because these evaluations are tied to specific methods, and because LIBERO, BadLIBERO, VAB-OmniGibson, EB-ALFRED, Safety-CHORES, and Bench2Drive differ in task goals, action spaces, and evidence levels, L1 still lacks a shared evaluation protocol.

\subsection{Multimodal Input and Interaction Security}
\label{sec:multimodal-input-interaction-security}

In large-model-driven embodied agents, multimodal input and interaction constitute the main interface between an agent and the external world. This stage concerns how an agent obtains and interprets information from environments and users. These inputs are not merely prompts passed to a language model; they may be transformed into scene interpretations, task representations, and model states that affect action generation. Input-side risks can therefore extend beyond semantic misunderstanding and influence downstream planning and physical execution.

The security boundary of L2 thus exceeds traditional prompt-level security. In end-to-end VLA systems, multimodal observations from environments and users directly affect visual-language-action generation. In LVLM-driven embodied agents, environmental text and scene semantics may also be interpreted as task cues or commands. Input risks may arise from malicious inputs, incomplete perception, or inconsistent observations. Once they enter the system, these risks may be amplified or misrouted during later processing and affect subsequent planning and execution.

\subsubsection{Attacks}

L2 attacks usually occur at runtime. They aim to change how an embodied agent interprets the environment or user intent through external inputs. Compared with textual attacks against ordinary LLMs, these attacks often have physical carriers and may further affect task planning and action generation. Existing methods can be summarized as sensor-level attacks, adversarial perception attacks, prompt injection, and jailbreak and action elicitation attacks. Table~\ref{tab:l2-attack-mechanisms} summarizes the representative attack mechanisms in this stage.

\begin{table}[t]
\centering
\caption{Attack mechanisms in multimodal input and interaction security}
\label{tab:l2-attack-mechanisms}
\scriptsize
\setlength{\tabcolsep}{3pt}
\begin{tabularx}{\dimexpr\textwidth-6pt\relax}{
L{0.17\textwidth}
L{0.20\textwidth}
Y
L{0.22\textwidth}
}
\toprule
\textbf{Attack Type} &
\textbf{Attack Methods} &
\textbf{Key Characteristics / Impact} &
\textbf{Datasets / Platforms} \tabularnewline
\midrule

Sensor-level attacks &
Phantom Menace~\cite{P1433} &
Injects physical camera or microphone signals; corrupted observations lead to misgrasp, drop, collision, or erratic execution. &
LIBERO-Spatial; LIBERO-Object; LIBERO-Goal; LIBERO-Long; Franka Panda. \tabularnewline
\midrule

Adversarial perception attacks &
UPA-RFAS~\cite{P5509} &
Uses a universal transferable patch; attention dominance and semantic mismatch cause cross-model task failure. &
BridgeData V2; LIBERO. \tabularnewline
\addlinespace[1.0ex]

&
ADVEDM~\cite{P5490} &
Removes or adds fine-grained object semantics; agents produce valid but task-incorrect embodied decisions. &
MS-COCO 2014; Dolphins Benchmark; DriveLM-nuScenes. \tabularnewline
\addlinespace[1.0ex]

&
EDPA~\cite{P5527} &
Disrupts visual-textual latent alignment; adversarial patches induce repeated incorrect actions. &
LIBERO-Spatial; LIBERO-Object; LIBERO-Goal; LIBERO-Long. \tabularnewline
\midrule

Prompt injection &
CHAI~\cite{P5541} &
Embeds deceptive text in the physical scene; environmental signs hijack the command layer. &
Known Images; Transferability Images; CARLA; AirSim; BARC. \tabularnewline
\addlinespace[1.0ex]

&
Mobile-robot prompt injection~\cite{P1743} &
Uses malicious or goal-hijacking instructions; multimodal prompts redirect navigation decisions. &
EyeSim VR; S4 bot. \tabularnewline
\midrule

Jailbreak and action elicitation attacks &
Textual VLA attacks~\cite{P5507} &
Uses token-level adversarial prompts; fixed or persistent target actions are elicited from VLA policies. &
LIBERO-Goal; LIBERO-Object; LIBERO-Spatial; LIBERO-10; Open X-Embodiment; HYDRA; SIMPLER. \tabularnewline
\addlinespace[1.0ex]

&
BADROBOT~\cite{P5483} &
Uses voice-based contextual jailbreaks; embodied LLM agents are induced to perform malicious physical actions. &
BADROBOT malicious physical action benchmark; PyBullet; RLBench; UR3e; myCobot 280-Pi. \tabularnewline

\bottomrule
\end{tabularx}
\end{table}

\textbf{Sensor-level attacks.} Sensor-level attacks act on the robot's physical sensing channel before inputs are represented as text, image features, or action-conditioned representations. The attacker does not need model or software access. physical signals distort sensor observations, causing later decisions to rely on corrupted environmental information. These attacks exploit the implicit trust between the sensor physical layer and the model input layer, where systems usually assume that observations faithfully reflect the environment state.

Phantom Menace~\cite{P1433} assumes that the attacker can inject physical signals only into a camera or microphone. It uses a Real-Sim-Real framework to collect real attack patterns, search attack parameters in simulation, and transfer them to a real robotic system. Experiments show that sensor-level attacks can cause misgrasp, drop, collision, and erratic execution behavior. The input security boundary of embodied agents therefore extends to the physical sensor interface, not only prompts or image files.

\textbf{Adversarial perception attacks.} Adversarial perception attacks use visual perturbations, physical patches, or cross-modal representation shifts to alter scene understanding. Their goal is not necessarily complete recognition failure, but a wrong yet superficially plausible visual-language interpretation that affects downstream decisions. This makes them difficult to detect through simple visual anomaly checks.

UPA-RFAS~\cite{P5509} constructs a universal transferable patch and improves cross-model, cross-task, and cross-view transferability through attention dominance and semantic mismatch. ADVEDM~\cite{P5490} removes target-object semantics or adds new object semantics, causing VLM-based embodied decision-making systems to produce logically coherent but task-incorrect decisions. EDPA~\cite{P5527} attacks VLAs through representation alignment by disrupting semantic consistency between visual latent representations and instruction representations, leading to visual misinterpretation and wrong actions.

\textbf{Prompt injection.} Prompt injection exploits confusion about text sources and semantic authority. External text is mistakenly interpreted as task instruction or control information. In embodied scenarios, the attacker need not access the user input interface; if the system can read environmental text, physical space itself becomes an indirect prompt-injection channel. The core risk is that the model may not reliably distinguish between observed environmental text and commands with execution authority.

CHAI~\cite{P5541} places deceptive natural-language instructions on readable signs in real scenes. It jointly optimizes textual content and visual presentation, causing LVLM-driven embodied systems to interpret environmental text as command-layer control information. Prompt-injection work on LLM-integrated mobile robots further shows that misleading text or multimodal input can enter the perception-brain-action pipeline and induce incorrect navigation instructions~\cite{P1743}.

\textbf{Jailbreak and action elicitation attacks.} Jailbreak and action elicitation attacks bypass safety constraints through interactive inputs and induce systems to generate dangerous plans or low-level actions. Unlike ordinary LLM jailbreaks, their target is not only harmful text generation, but also outputs that affect system behavior.

Textual attacks against robotic VLAs adapt jailbreak methods to the action space by using token-level textual attacks to induce fixed actions or action sequences~\cite{P5507}. BADROBOT~\cite{P5483} attacks embodied LLM systems through voice-based interactions, using contextual jailbreak, safety misalignment, and conceptual deception to induce physically malicious behavior.

\subsubsection{Defenses}

Existing defenses intervene at different points of the input-processing chain to reduce the influence of external inputs on downstream planning and action generation. According to their defense target and primary intervention point, they can be summarized as input moderation and jailbreak detection, representation sanitization, perception-aware robustness, and interaction-level guardrails. Table~\ref{tab:l2-defense-mechanisms} summarizes the representative defense mechanisms in this stage.

\begin{table}[t]
\centering
\caption{Defense mechanisms in multimodal input and interaction security}
\label{tab:l2-defense-mechanisms}
\scriptsize
\setlength{\tabcolsep}{3pt}
\begin{tabularx}{\dimexpr\textwidth-6pt\relax}{
L{0.26\textwidth}
L{0.18\textwidth}
L{0.22\textwidth}
Y
}
\toprule
\textbf{Defense Type} &
\textbf{Defense Methods} &
\textbf{Verification Scenario} &
\textbf{Datasets / Platforms} \tabularnewline
\midrule

Input moderation and jailbreak detection &
Pinpoint~\cite{P4845} &
Non-embodied evaluation &
EAsafetyBench. \tabularnewline
\addlinespace[1.0ex]

&
J-DAPT~\cite{P1306} &
Non-embodied evaluation &
DAQUAR; JB28K; LingoQA; nuScenes; ABOships-PLUS; LaRS. \tabularnewline
\midrule

Representation sanitization &
SAFE-Dict~\cite{P1415} &
Non-embodied evaluation &
Libero-Harm; BadRobot; RoboPair; IS-Bench. \tabularnewline
\midrule

Perception-aware robustness &
Safe-Night VLA~\cite{P1403} &
Real-robot platform &
Franka Emika Panda. \tabularnewline
\addlinespace[1.0ex]

&
TrustNavGPT~\cite{P1826} &
Real-robot platform &
DNIA; RoboTHOR; LoCoBot. \tabularnewline
\midrule

Interaction-level guardrails &
M-CoDAL~\cite{P219} &
Real-robot platform &
M-CoDAL safety dataset; Hello Robot Stretch. \tabularnewline
\addlinespace[1.0ex]

&
Unified security-safety framework~\cite{P5073} &
Real-robot platform &
EyeSim VR; Pioneer mobile robot. \tabularnewline

\bottomrule
\end{tabularx}
\end{table}

\textbf{Input moderation and jailbreak detection.} Input moderation and jailbreak detection identify malicious intent before user instructions or multimodal contexts enter downstream reasoning. They are suitable for explicit malicious instructions, disguised requests, and jailbreak inputs. Their main limitation is that detection usually operates on text or multimodal representations and cannot directly defend against physical sensor-layer attacks.

Pinpoint~\cite{P4845} addresses input moderation for embodied agents and observes that functional prompts can interfere with conventional moderation. It uses masked attention in intermediate layers of embodied LLMs, marks user-instruction boundaries with special tokens, and extracts instruction-related hidden states for malicious-instruction moderation. J-DAPT~\cite{P1306} formulates robotic jailbreak detection as a cross-domain problem. It transfers general jailbreak-detection priors to robotic scenarios and combining multimodal fusion with domain adaptation to improve stability.

\textbf{Representation sanitization.} Representation sanitization intervenes after raw inputs are encoded and fused. Unlike filters that inspect only text or images, these methods identify and weaken high-risk semantics in internal representations. They are suitable for unsafe intents already in latent space, but depend on whether dangerous concepts can be stably decomposed and localized.

SAFE-Dict~\cite{P1415} deploys defense at the fused latent representation layer of a VLA. It learns a concept dictionary, decomposes hidden activations into interpretable concept coefficients, computes harmfulness scores, weakens high-risk coefficients, and reconstructs sanitized latents for action generation. This avoids retraining the whole VLA and can act after multimodal fusion. However, it cannot verify the source of environmental text or directly block physical sensor attacks such as camera laser or microphone spoofing.

\textbf{Perception-aware robustness.} Perception-aware robustness improves the quality and reliability of input information rather than only detecting malicious content. This direction is useful for perceptual blind spots, speech ambiguity, and uncertainty in user instructions. It asks whether the system has enough information to support safe decisions.

Safe-Night VLA~\cite{P1403} integrates LWIR thermal perception into a frozen visual-language backbone. Thermal information helps cover physical properties missing from RGB input, while control barrier functions mitigate execution risks caused by perceptual blind spots. TrustNavGPT~\cite{P1826} targets audio-guided robot navigation and processes speech transcription with vocal cue features to estimate the certainty and trustworthiness of navigation instructions.

\textbf{Interaction-level guardrails.} Interaction-level guardrails build safety boundaries among user intent, environmental context, and system responses, structuring the link between input understanding and later execution. These mechanisms usually depend on task-specific rules and interaction settings, making them more suitable as local guardrails than general input defenses.

M-CoDAL~\cite{P219} targets safety-critical multimodal dialogue in assistive robots. It uses multimodal safety data, discourse coherence relations, and active learning to train the dialogue system, and evaluates it through user studies on Hello Robot Stretch. The unified security-safety framework for LLM-integrated mobile robots~\cite{P5073} combines prompt assembly, state management, and safety validation, enabling inputs to be structurally parsed and checked before action generation.

\subsubsection{Evaluation}

L2 evaluation asks whether multimodal inputs change an agent's perception, task understanding, or downstream behavior. Existing evaluations cover four main objects: visual and instruction perturbations, natural-language red teaming, physical environmental changes, and hazardous-instruction diagnosis.

VLA-RISK~\cite{P5511} evaluates embodied decision risks under visual and instruction perturbations using LIBERO, VLABench, and nuScenes. Embodied Red Teaming~\cite{P5523} generates diverse instructions from environmental images and task descriptions, and tests the robustness of language-conditioned robots on CALVIN, RLBench, and SimplerEnv. These benchmarks examine whether agents preserve correct task understanding and execution tendencies under input variation.

Physical feasibility is also important. Eva-VLA~\cite{P5525} parameterizes object pose, lighting changes, and adversarial patch placement, uses black-box optimization to search for worst-case scenarios, and evaluates VLA physical robustness in LIBERO and real robotic-arm settings. AGENTSAFE~\cite{P5486} constructs SAFE-THOR, SAFE-VERSE, and SAFE-DIAGNOSE to diagnose whether embodied VLM agents facing hazardous instructions can translate danger recognition into safe planning and execution. These evaluations not only measure whether input perturbations cause task failure, but also help locate how errors propagate across perception, planning, and execution.

\subsection{Semantic Reasoning, Risk Awareness, and Task Planning Security}
\label{sec:semantic-reasoning-task-planning-security}

In large-model-driven embodied agents, semantic reasoning, risk awareness, and task planning form the transition from input understanding to action execution. At this stage, user goals, environmental information, and system states are transformed into executable task plans. Compared with multimodal input security, L3 no longer focuses only on what information the system receives, but on how the system understands goals and forms executable plans.

The security boundary of L3 therefore exceeds ordinary textual safety in LLMs. Planning feasibility and planning safety must be distinguished: a plan may satisfy a formal task objective while violating safety constraints, physical conditions, or domain norms. In embodied systems, planning output is usually not the final result, but upstream input for action generation and control execution. Once an unsafe plan is adopted downstream, a semantic-level error can become dangerous behavior.

\subsubsection{Attacks}

L3 attacks aim to manipulate the model's task-planning and decision-making process, causing it to generate action plans aligned with the attacker's intent. Attack success depends not only on whether the model generates dangerous text, but also on whether the output can be parsed and executed by downstream tools or action modules. Existing attacks mainly include decision-level adversarial prompting and policy-executable jailbreak attacks. Table~\ref{tab:l3-attack-mechanisms} summarizes the representative attack mechanisms in this stage.

\begin{table}[t]
\centering
\caption{Attack mechanisms in semantic reasoning, risk awareness, and task planning security}
\label{tab:l3-attack-mechanisms}
\scriptsize
\setlength{\tabcolsep}{3pt}
\begin{tabularx}{\dimexpr\textwidth-6pt\relax}{
L{0.20\textwidth}
L{0.14\textwidth}
Y
L{0.25\textwidth}
}
\toprule
\textbf{Attack Type} &
\textbf{Attack Methods} &
\textbf{Key Characteristics / Impact} &
\textbf{Datasets / Platforms} \tabularnewline
\midrule

Decision-level adversarial prompting &
EIRAD~\cite{P5489} &
Uses adversarial suffixes to perturb multi-step task planning; generated plans deviate toward irrelevant or malicious goals. &
EIRAD; AI2-THOR. \tabularnewline
\midrule

Policy-executable jailbreak attacks &
POEX~\cite{P5498} &
Optimizes word-level adversarial suffixes toward harmful executable policies; physical damage becomes the attack objective. &
Harmful-RLBench; RLBench; CoppeliaSim; Franka Panda; Unitree G1. \tabularnewline
\addlinespace[1.0ex]

&
ROBOPAIR~\cite{P5506} &
Adapts chatbot jailbreaks to robot action interfaces; harmful prompts are constrained by robotic APIs and execution syntax. &
Dolphins harmful-action set; Clearpath Jackal harmful-action task set; Unitree Go2 harmful-action task set. \tabularnewline

\bottomrule
\end{tabularx}
\end{table}

\textbf{Decision-level adversarial prompting.} Decision-level adversarial prompting uses adversarial prompts or suffixes to affect high-level task decomposition, causing generated plan steps to deviate from the original goal or align with the attacker's goal. The attack target is the planning output itself, rather than the safety of a single input token. For embodied agents, generated steps may be parsed by execution modules as robotic task procedures, so high-level plan deviation can affect overall system behavior.

EIRAD~\cite{P5489} constructs a decision-level robustness benchmark for LLM-based embodied models. It adapts the GCG attack process to multi-step task-plan generation, using target-task keywords to initialize adversarial suffixes and step slicing with similarity calculation to test whether generated steps match the target attack task. This directly interferes with task decomposition and causes deviation from user intent or safety goals at the planning stage.

\textbf{Policy-executable jailbreak attacks.} Policy-executable jailbreak attacks shift jailbreak goals from harmful text to executable policies. The attacker must bypass safety alignment while making outputs acceptable to program interfaces or action modules. The result is both harmful and executable, directly connecting semantic safety problems to action-execution risks.

POEX~\cite{P5498} focuses on harmful yet executable robot policies. It uses hidden-layer gradient optimization to make adversarial suffixes bypass safety alignment while improving policy executability. ROBOPAIR~\cite{P5506} adapts PAIR-style chatbot jailbreaks to robot action interfaces, using robot-specific system prompts and a syntax checker so that generated jailbreak prompts satisfy robotic API syntax requirements.

\subsubsection{Defenses}

L3 defense mechanisms mainly act on task planning and the critical steps before execution. They have expanded from prompt constraints to formal constraint and pre-execution verification, runtime safety logic, risk reasoning and deliberation, and counterfactual self-correction. Table~\ref{tab:l3-defense-mechanisms} summarizes the representative defense mechanisms in this stage.

\begin{table}[t]
\centering
\caption{Defense mechanisms in semantic reasoning, risk awareness, and task planning security}
\label{tab:l3-defense-mechanisms}
\scriptsize
\setlength{\tabcolsep}{3pt}
\begin{tabularx}{\dimexpr\textwidth-6pt\relax}{
L{0.26\textwidth}
L{0.18\textwidth}
L{0.22\textwidth}
Y
}
\toprule
\textbf{Defense Type} &
\textbf{Defense Methods} &
\textbf{Verification Scenario} &
\textbf{Datasets / Platforms} \tabularnewline
\midrule

Formal constraint and pre-execution verification &
SafeGate~\cite{P1280} &
Simulation environment &
SafeGate 230-task benchmark; AI2-THOR. \tabularnewline
\addlinespace[1.0ex]

&
Safety Chip~\cite{P5242} &
Real-robot platform &
VirtualHome; Boston Dynamics Spot. \tabularnewline
\addlinespace[1.0ex]

&
Cross-layer sequence supervision~\cite{P5524} &
Real-robot platform &
VirtualHome; Gazebo; Franka Panda; TurtleBot. \tabularnewline
\midrule

Runtime safety logic &
RoboSafe~\cite{P5542} &
Real-robot platform &
SafeAgentBench; AI2-THOR; myCobot 280-Pi. \tabularnewline
\midrule

Risk reasoning and deliberation &
MADRA~\cite{P5482} &
Simulation environment &
SafeAware-VH; SafeAgentEnv; VirtualHome; AI2-THOR; ALFRED. \tabularnewline
\midrule

Counterfactual self-correction &
Counterfactual VLA~\cite{P1421} &
Non-embodied evaluation &
Dtraj; Dmeta; DCF. \tabularnewline

\bottomrule
\end{tabularx}
\end{table}

\textbf{Formal constraint and pre-execution verification.} Formal constraint and pre-execution verification translate natural-language tasks or safety requirements into checkable formal constraints, and verify plans before execution. These methods provide explicit inspection objects and failure reasons. They are suitable for tasks with clear rules, representable states, and formalizable constraints. Their effect depends on the completeness of constraint expressions and the reliability of environment observation and symbolization.

SafeGate~\cite{P1280} performs safety authorization before natural-language tasks enter planning and code generation. It converts tasks into safety contracts and uses steps such as Z3 SMT-based Plan Verification to decide whether to authorize, defer, or reject a task. Safety Chip~\cite{P5242} maps user natural-language safety constraints into Linear Temporal Logic formulas, and inserts them as a queryable safety constraint module into an LLM agent. The system can replan according to LTL constraints in VirtualHome and real-robot experiments. Formal interfaces clarify planning safety constraints. They also require the system to accurately map natural-language risks, environment states, and action consequences into symbolic expressions.

Cross-layer sequence supervision~\cite{P5524} further converts LTL safety constraints into nondeterministic B"{u}chi automata and constructs a cross-layer safety supervisor. The supervisor checks LLM-generated action sequences at the task-planning layer and passes unsafe-action-related regions to the motion-planning layer as virtual obstacles. This extends task-plan constraints into motion planning, but still depends on predefined constraints and accurate environmental grounding.

\textbf{Runtime safety logic.} Runtime safety logic uses executable predicates, safety memory, or contextual rules to continuously check whether plans and actions satisfy safety conditions before and during execution. Compared with one-time pre-execution verification, these methods better address implicit hazards and temporally dependent risks. Their key premise is that safety predicates are generated accurately and safety memory covers the current context.

RoboSafe~\cite{P5542} targets VLM-driven embodied agents and proposes executable predicate-based safety logic as a runtime guardrail. It combines Hybrid Long-Short Safety Memory, Backward Reflective Reasoning, and Forward Predictive Reasoning to make BLOCK or PASS decisions before the next action is executed.

\textbf{Risk reasoning and deliberation.} Risk reasoning and deliberation use external safety knowledge, multi-agent assessment, or structured reasoning to judge task and plan safety. Compared with formal constraints, these mechanisms are more flexible and can cover safety situations that are difficult to symbolize. However, their reasoning quality depends on knowledge-base coverage, evaluator reliability, and model-generated content, so they cannot provide strict safety guarantees.

SafeMindAgent~\cite{P5487} uses the four-stage risk model in SafeMindBench, dividing the embodied-agent reasoning chain into task understanding, environment perception, high-level planning, and low-level action generation. It integrates an external safety constraint knowledge base with cascaded verification to check and revise plans during reasoning. MADRA~\cite{P5482} adopts training-free Multi-Agent Debate Risk Assessment, where multiple LLM-based risk assessment agents make structured judgments and an evaluator with consensus voting produces the final decision. These methods can improve risk identification and task screening before planning, but do not replace formal verification or runtime control protection.

\textbf{Counterfactual self-correction.} Counterfactual self-correction asks the model to infer possible consequences of its current plan before generating a final trajectory or decision, and then revise unsafe or suboptimal plans. It is suitable for prospective risk judgment, especially in autonomous driving and dynamic-environment decision-making. Its effect depends on the accuracy of counterfactual reasoning and whether training or annotation data cover key risk scenarios.

Counterfactual VLA~\cite{P1421} targets autonomous driving VLAs. It proposes a self-reflection chain of meta-actions $\rightarrow$ counterfactual reasoning $\rightarrow$ updated meta-actions $\rightarrow$ trajectory. The method uses time-segmented meta-actions to align language and actions, performs counterfactual reasoning based on visual context and proposed actions, and revises unsafe plans before generating the final trajectory. This provides an internal self-check at the planning stage, but current validation mainly relies on autonomous-driving data and trajectory-level metrics, leaving its generalization to broader robotic tasks open.

\subsubsection{Evaluation}

Evaluation at this stage mainly asks whether a model can transform task goals, environment states, safety constraints, and action requirements into safe plans. Existing evaluation mechanisms broadly cover planning safety, reasoning-chain localization, interactive task execution, dynamic-environment decision-making, and high-risk domain testing.

SAFEL and EMBODYGUARD~\cite{P5502} use Command Refusal Test and Plan Safety Test to evaluate explicit danger refusal and implicit hazard planning, and localize failures to goal interpretation, state-transition modeling, action ordering, and related steps. DESPITE~\cite{P1286} uses deterministic formal validation to judge whether an action plan is feasible and safe, distinguishing planning capability from safety awareness through Feasibility, Safety, Safety Precision, and Safety Intention. SafeMindBench~\cite{P5487} maps safety problems to a multi-stage reasoning chain, enabling analysis of failure locations during task understanding, environment perception, and plan generation.

Interactive and closed-loop evaluations examine whether planning decisions affect later execution outcomes. SafeAgentBench~\cite{P5499} evaluates embodied LLM agents on hazardous and safe tasks in AI2-THOR-based SafeAgentEnv. The HAZARD challenge~\cite{P5133} simulates dynamic disaster environments in ThreeDWorld and evaluates rescue decisions using Value, Step, and Damage. PADriver~\cite{P4190} provides a Highway-Env-based closed-loop autonomous-driving evaluation, examining MLLM performance in scene understanding, hazard-level estimation, and action decision-making.

Semantic safety and domain-risk evaluations examine whether the model adjusts behavior according to task semantics and scenario-specific norms. HazardArena~\cite{P1383} uses safe/unsafe twin scenarios. It changes semantic context while keeping action requirements the same, and evaluates whether VLAs truly understand semantic risk. The Robotic Health Attendant safety benchmark~\cite{P1281} targets medical robot control. It uses harmful instructions and paired benign instructions to evaluate whether LLMs follow medical ethics and patient-safety constraints.

\subsection{Action Execution, Control Chain, and Physical Feedback Security}
\label{sec:action-execution-control-chain-physical-feedback-security}

Action execution, control chain, and physical feedback form the final transition from semantic plans to the real environment. High-level task goals and planning results are translated into concrete action sequences and control signals at this stage, and these signals form a closed loop with environmental feedback. This stage directly concerns physical behavior. Its safety depends on whether actions satisfy temporal constraints, whether control commands are filtered for safety, whether execution can be monitored, and whether environmental feedback can trigger necessary safety responses.

\subsubsection{Attacks}

L4 attacks directly target the action space and execution chain. These attacks usually decompose a malicious goal into a set of superficially reasonable actions. Each individual action may appear acceptable, but their combined physical consequence can be risky. The representative direction in existing work is action-level manipulation attacks. Table~\ref{tab:l4-attack-mechanisms} summarizes the representative attack mechanisms in this stage.

\begin{table}[t]
\centering
\caption{Attack mechanisms in action execution, control chain, and physical feedback security}
\label{tab:l4-attack-mechanisms}
\scriptsize
\setlength{\tabcolsep}{3pt}
\begin{tabularx}{\dimexpr\textwidth-6pt\relax}{
L{0.17\textwidth}
L{0.14\textwidth}
Y
L{0.20\textwidth}
}
\toprule
\textbf{Attack Type} &
\textbf{Attack Methods} &
\textbf{Key Characteristics / Impact} &
\textbf{Datasets / Platforms} \tabularnewline
\midrule

Action-level manipulation attacks &
Blindfold~\cite{P116} &
Decomposes malicious intent into benign-looking action sequences; cover actions conceal harmful physical consequences while preserving executability. &
BadRobot; SafeAgentBench; VirtualHome; Habitat; ManiSkill; RoboTHOR; UFactory xArm 6. \tabularnewline

\bottomrule
\end{tabularx}
\end{table}

\textbf{Action-level manipulation attacks.} Action-level manipulation attacks exploit the gap between semantic safety checks and physical consequences. Semantic safeguards often check whether user requests, model responses, or high-level plans contain explicit malicious intent. However, risks in action sequences may arise from combinations of ordinary actions: a single step may not look dangerous, while continuous execution may damage devices or cause other physical harms.

Blindfold's Adversarial Proxy Planning~\cite{P116} uses a local open-source LLM as an adversarial planner to generate candidate action sequences from malicious intent and environmental context. An intent obfuscator then inserts cover actions to reduce the visibility of malicious intent, while a rule-based verifier checks symbolic constraints to ensure executability. This hides dangerous intent inside action-level compositions. Execution-layer risk therefore no longer comes from a single explicit dangerous action, but from cumulative physical consequences across time steps.

\subsubsection{Defenses}

L4 defenses directly act on action generation and execution. They impose constraints before actions enter the physical environment and during execution. The main methods include runtime control shielding, constrained action decoding, execution monitoring and feedback, action hallucination mitigation, and domain-specific compliance control. Table~\ref{tab:l4-defense-mechanisms} summarizes the representative defense mechanisms in this stage.

\begin{table}[t]
\centering
\caption{Defense mechanisms in action execution, control chain, and physical feedback security}
\label{tab:l4-defense-mechanisms}
\scriptsize
\setlength{\tabcolsep}{3pt}
\begin{tabularx}{\dimexpr\textwidth-6pt\relax}{
L{0.26\textwidth}
L{0.18\textwidth}
L{0.22\textwidth}
Y
}
\toprule
\textbf{Defense Type} &
\textbf{Defense Methods} &
\textbf{Verification Scenario} &
\textbf{Datasets / Platforms} \tabularnewline
\midrule

Runtime control shielding &
Language-conditioned MPC safety filtering~\cite{P1311} &
Real-robot platform &
Habitat 3.0; TurtleBot 2. \tabularnewline
\addlinespace[1.0ex]

&
AEGIS~\cite{P1434} &
Simulation environment &
SafeLIBERO; LIBERO-Spatial; LIBERO-Object; LIBERO-Goal; LIBERO-Long. \tabularnewline
\addlinespace[1.0ex]

&
LLM-DiSC~\cite{P2697} &
Simulation environment &
Polygonal-obstacle Multi-robot simulation platform. \tabularnewline
\midrule

Constrained action decoding &
SafeDec~\cite{P1689} &
Simulation environment &
CHORES; SafetyCHORES; AI2-THOR. \tabularnewline
\midrule

Execution monitoring and feedback &
RoboCritics~\cite{P182} &
Real-robot platform &
UR3e. \tabularnewline
\addlinespace[1.0ex]

&
SAFE~\cite{P5543} &
Real-robot platform &
LIBERO-10; SimplerEnv; Franka Emika Panda; WidowX 250. \tabularnewline
\addlinespace[1.0ex]

&
VLA uncertainty monitoring~\cite{P1397} &
Non-embodied evaluation &
LIBERO-Spatial; LIBERO-Object; LIBERO-Goal; LIBERO-10. \tabularnewline
\midrule

Action hallucination mitigation &
EOCD~\cite{P4996} &
Simulation environment &
LIBERO-Spatial; LIBERO-Object; LIBERO-Goal; LIBERO-Long. \tabularnewline
\midrule

Domain-specific compliance control &
RoboChemist~\cite{P5461} &
Real-robot platform &
Cobot Magic ALOHA. \tabularnewline

\bottomrule
\end{tabularx}
\end{table}

\textbf{Runtime control shielding.} Runtime control shielding intervenes after action generation and before low-level execution. It uses control filters, safety barriers, or optimizers to revise nominal actions, so that the robot satisfies physical constraints while preserving task intent as much as possible. These methods mainly constrain execution-time physical risks, and their effectiveness depends on environment modeling and control accuracy.

Language-conditioned MPC safety filtering~\cite{P1311} translates natural-language safety requirements into structured specifications, grounds relevant objects and regions in a three-dimensional environment, and uses an MPC-based safety filter to revise control inputs online. AEGIS~\cite{P1434} adds a plug-and-play safety constraint layer after an existing VLA model. It uses a VLM to identify task-relevant hazardous objects and a CBF-QP solver to minimally modify the VLA nominal action for collision avoidance. These methods translate language constraints or visual grounding into execution-layer control constraints.

Control shielding in multi-robot settings must also handle interaction risks. LLM-DiSC~\cite{P2697} combines an LLM-generated reference planner with a CBF-based modifier for distributed multi-robot coordination. The modifier revises control strategies according to inter-robot and robot-obstacle safety distances, and returns simulation feedback to the LLM for iterative optimization. In multi-robot execution, safety control constrains not only individual trajectories, but also coordination stability among robots.

\textbf{Constrained action decoding.} Constrained action decoding embeds safety constraints directly into action generation, rather than filtering unsafe actions after generation. Autoregressive robotic policies predict future actions step by step, so unsafe behavior can accumulate during sampling. Introducing safety constraints into the decoding layer allows the system to filter or reweight candidate actions before rollout, reducing unsafe action sampling.

SafeDec~\cite{P1689} introduces STL safety specifications into the decoding layer of autoregressive robot navigation policies. It proposes Hard Constrained Decoding and Robustness Constrained Decoding. The method uses an approximate dynamics model to predict future states for candidate actions, and masks or reweights them according to STL satisfaction or robustness. This allows formal temporal safety requirements to be incorporated directly into the action decoding process.

\textbf{Execution monitoring and feedback.} Execution monitoring and feedback concern whether abnormalities can be detected during execution and whether the system can provide useful intervention signals.

RoboCritics~\cite{P182} integrates expert-informed critics into an LLM-based robot programming workflow. It collects motion-level traces in simulation and UR3e physical execution, checks violations such as collisions, joint speeds, and end-effector pose, and provides structured feedback and one-click fixes for program revision. SAFE~\cite{P5543} extracts representations from the last internal layer of a VLA, uses SAFE-MLP or SAFE-LSTM to predict failure scores, and calibrates thresholds with functional conformal prediction. It evaluates execution failure detection on OpenVLA, $\pi_0$, and $\pi_0$-FAST.

Uncertainty monitoring is another important signal source. Related work notes that global average entropy may hide risk peaks in critical time windows, and proposes Sliding Window Pooling, Action Transfer Reweighting, and DoF-Adaptive Calibration to capture brief but critical uncertainty spikes~\cite{P1397}. These methods can support replanning or human takeover, but unified recovery procedures after abnormality detection remain underdeveloped.

\textbf{Action hallucination mitigation.} Action hallucination mitigation addresses action outputs inconsistent with visual observations. This problem may originate from upstream perception bias, but at L4 it appears as wrong actions or targets. Mitigation requires reducing hallucination-prone actions while preserving multiple plausible execution paths.

EOCD~\cite{P4996} is a training-free decoding framework for suppressing VLA action-level hallucination. It uses Decision Path Mining via Contrastive Decoding to compare action distributions under original and perturbed visual inputs, and Entropy-Maximized Contrastive Optimization to adjust decoding hyperparameters. This avoids excessive concentration of probability mass on a single path and reduces action hallucination caused by visual misjudgment.

\textbf{Domain-specific compliance control.} Domain-specific compliance control examines execution safety under high-risk task norms. Its scope is narrower, but it highlights the difference between task completion and compliant execution in real applications.

RoboChemist~\cite{P5461} targets chemical experiments. It uses a VLM for planning and monitoring and a VLA to execute atomic chemical operations. The system annotates grasping points or target regions with visual prompts. An outer VLM monitor judges subtask completion and repeats the current primitive task when necessary. Success Rate and Compliance Rate jointly evaluate task completion and compliance with experimental norms. In high-risk scenarios such as chemical experiments, execution safety must incorporate domain norms rather than rely only on task completion rate.

\subsubsection{Evaluation}

Existing work evaluates two broad objects: whether execution trajectories satisfy temporal safety properties, and whether action quality can be assessed at fine granularity.

SAFEMANIP~\cite{P1382} uses LTLf safety property templates to monitor finite traces of manipulation rollouts, mapping execution processes into symbolic predicate traces. The benchmark covers RoboCasa365 household tasks and multiple VLA policies. It reports both task completion and temporal safety, revealing collisions, unstable releases, cross-contamination, and other problems that may appear even in successful tasks. This mechanism distinguishes task completion from execution safety, with evidence mainly from simulated trajectories and state information.

Eval-Actions and AutoEval~\cite{P1410} focus on action quality and execution-source authenticity. Because binary success rates can hide jerky execution, unsafe motion, and source ambiguity, they construct datasets containing teleoperation and policy-generated trajectories. They train automatic evaluators with Expert Grading, Rank-Guided preferences, and Chain-of-Thought annotations. The evaluation target is therefore expanded from task completion to whether execution is smooth, safe, and trustworthy.

\subsection{Deployment, Multi-agent Ecosystems, and Human-centered Impacts}
\label{sec:deployment-multi-agent-human-centered-impacts}

Once large-model-driven embodied agents enter real operational environments, their security and privacy boundaries are no longer limited to a single model or interaction. They extend to the broader deployment environment formed by cloud services, edge devices, and multi-stakeholder collaboration. The core issue at this stage is how embodied agents maintain trustworthy communication, controllable permissions, governable privacy, and traceable accountability during long-term operation, multi-party collaboration, and real-user participation.

Risks at this stage have clear system-level and time-accumulative characteristics. Communication-link tampering may directly change robot control commands. Collaborative messages may propagate malicious intent in multi-robot systems. Long-term edge perception may accumulate sensitive behavioral profiles. Human users may also over-authorize or mistakenly rely on a system because of trust mismatch. Deployment-stage security therefore needs to be understood from the overall mode of system operation, including structures and interaction mechanisms beyond the model itself. Its boundary covers cloud-edge communication, multi-agent collaboration, communication channels, authorization mechanisms, and human users. In human-robot coexistence environments, users are both interaction subjects and important bearers of risk and responsibility.

\subsubsection{Attacks}

Deployment-stage attacks mainly target communication links, collaboration protocols, and human-robot interaction after deployment. These attacks usually do not require modifying model weights, and they may not directly manipulate model inputs. Instead, attackers often influence system decisions, physical execution, and user cognition indirectly by manipulating information flows or interaction environments during system operation. Existing directions can be summarized as communication hijacking attacks, multi-agent propagation attacks, and trust-manipulation attacks. Table~\ref{tab:l5-attack-mechanisms} summarizes the representative attack mechanisms in this stage.

\begin{table}[t]
\centering
\caption{Attack mechanisms in deployment, multi-agent ecosystems, and human-centered impacts}
\label{tab:l5-attack-mechanisms}
\scriptsize
\setlength{\tabcolsep}{3pt}
\begin{tabularx}{\dimexpr\textwidth-6pt\relax}{
L{0.22\textwidth}
L{0.16\textwidth}
Y
L{0.20\textwidth}
}
\toprule
\textbf{Attack Type} &
\textbf{Attack Methods} &
\textbf{Key Characteristics / Impact} &
\textbf{Datasets / Platforms} \tabularnewline
\midrule

Communication hijacking attacks &
From Prompts to Motors~\cite{P5505} &
Manipulates JSON messages in the robot--LLM communication link; bypasses obstacle avoidance, overrides stop commands, and deceives user feedback. &
Commercial vacuum robot; Raspberry Pi 5. \tabularnewline
\midrule

Multi-agent propagation attacks &
InfectBot~\cite{P1275} &
Compromises one entry robot through dialogue; malicious intent spreads through peer communication and induces coordinated unsafe behavior. &
NVIDIA Isaac Sim; Isaac Lab; ROS 2 Humble; Unitree ROS 2. \tabularnewline
\midrule

Trust-manipulation attacks &
Trust-aware HRI attack~\cite{P3593} &
Uses environment perturbations to distort trust-aware decisions; collaboration efficiency and human trust decrease. &
Unreal Engine; quadruped robot dog. \tabularnewline

\bottomrule
\end{tabularx}
\end{table}

\textbf{Communication hijacking attacks.} Communication hijacking attacks manipulate message flows between deployed embodied agents and cloud models or edge devices. Attackers exploit verification gaps in communication links and tamper with structured messages, causing robots to execute wrong control commands or ignore safety constraints, with possible device damage, environmental harm, or direct harm to humans.

From Prompts to Motors~\cite{P5505} proposes a gray-box MITM attack against an LLM-enabled vacuum robot. The attacker has Wi-Fi credentials and API endpoint knowledge, but cannot access the robot's internal code or real-time state. Using ARP spoofing, the attacker intercepts JSON messages between the robot and the ChatGPT API, and performs indirect prompt injection and output manipulation. By tampering with obstacle-detection information, the attacker can bypass pet avoidance. The attacker can also change \texttt{stop\_cleaning()} into \texttt{start\_cleaning()} or \texttt{continue\_cleaning()}, causing operation against the user's stop command. In addition, tampered status feedback may cause the user to misjudge system state and task outcome, delaying intervention and weakening system trustworthiness.

\textbf{Multi-agent propagation attacks.} Multi-agent propagation attacks use collaborative messages to spread malicious intent among agents. A local compromise may be amplified through internal communication and task collaboration into a system-level event. The security of a multi-agent system depends not only on each agent's refusal capability, but also on whether the collaboration protocol verifies message sources, limits propagation privileges, and blocks malicious forwarding.

InfectBot~\cite{P1275} treats inter-robot communication as the primary attack surface. The attacker interacts with only one entry robot through natural language. A prompt sequence induces this robot to forward a malicious protocol, which then spreads unsafe intent through peer communication. After a single entry node is manipulated, malicious influence may diffuse along collaborative communication links.

\textbf{Trust-manipulation attacks.} Trust-manipulation attacks do not directly modify control commands. Instead, they manipulate environmental or system states that affect human trust judgments, indirectly changing human-robot collaboration outcomes. In trust-aware HRI systems, human trust is often an input to robot decisions. If an attacker affects trust estimation, they may alter collaboration rhythm and human takeover behavior.

Existing work models trust-aware human-robot collaboration as a sequential decision process. Under a closed-box assumption, it constructs a surrogate model and uses FGSM to generate environmental perturbations that reduce both collaboration efficiency and human trust~\cite{P3593}. In human-robot collaboration scenarios, the trust model itself therefore needs to be treated as a protected security object.

\subsubsection{Defenses}

Defenses at this stage require system-level design and place greater emphasis on system-level control and governance mechanisms. The main methods include task orchestration and provenance, access control and mission recovery, optimization-mediated multi-agent coordination, and privacy and HRI governance. Table~\ref{tab:l5-defense-mechanisms} summarizes the representative defense mechanisms in this stage.

\begin{table}[t]
\centering
\caption{Defense mechanisms in deployment, multi-agent ecosystems, and human-centered impacts}
\label{tab:l5-defense-mechanisms}
\scriptsize
\setlength{\tabcolsep}{3pt}
\begin{tabularx}{\dimexpr\textwidth-6pt\relax}{
L{0.27\textwidth}
L{0.18\textwidth}
L{0.21\textwidth}
Y
}
\toprule
\textbf{Defense Type} &
\textbf{Defense Methods} &
\textbf{Verification Scenario} &
\textbf{Datasets / Platforms} \tabularnewline
\midrule

Task orchestration and provenance &
Decentralized task planner~\cite{P1324} &
Non-embodied evaluation &
SkillChain-RTD. \tabularnewline
\midrule

Access control and mission recovery &
FlyAdapt~\cite{P2230} &
Simulation environment &
DJI Matrice 300 RTK; PX4; AirSim; NVIDIA Jetson AGX Xavier. \tabularnewline
\addlinespace[1.0ex]

&
DER~\cite{P2497} &
Simulation environment &
Gazebo; PX4; ROS Noetic. \tabularnewline
\midrule

Optimization-mediated multi-agent coordination &
Hierarchical LLMs In-the-Loop Optimization~\cite{P4762} &
Real-robot platform &
Multi-robot target-tracking simulation; aerial drones; ground robots. \tabularnewline
\midrule

Privacy and HRI governance &
MALLM~\cite{P4343} &
Long-term deployment &
NVIDIA Jetson NX; four-person student apartment; Intel RealSense D435i; TI IWR6843; FLIR Lepton 3.5; Sense Energy Monitor. \tabularnewline

\bottomrule
\end{tabularx}
\end{table}

\textbf{Task orchestration and provenance.} Task orchestration and provenance mechanisms address task decomposition, order verification, and result auditability in multi-robot or multi-organization environments. Their goal is to preserve source traceability and accountability boundaries as task plans pass among multiple models, agents, or services. For systems relying on multiple LLM oracles or collaborative agents, these mechanisms improve transparency and consistency in task orchestration.

The Decentralized Intent-Based Multi-Robot Task Planner~\cite{P1324} records user request hashes and LLM oracle responses through an Oracle Smart Contract. It maintains a robot registry, assigns validated plan segments through a Planner Smart Contract, and aggregates outputs from multiple LLM oracles using Longest Common Subsequence and historical reputation. The method mainly targets multi-robot task decomposition and reduces orchestration risk through task provenance and sequential consistency.

\textbf{Access control and mission recovery.} Access control and mission recovery mechanisms address permission management, tool-call constraints, and anomaly recovery in deployed systems. For UAV and cloud-edge embodied systems, dynamically changing tasks and node states make static access control difficult. These defenses therefore adjust permissions at runtime and combine state verification with mission recovery to maintain continuity and safety after anomalies or attacks.

FlyAdapt~\cite{P2230} targets real-time dynamic access control in multi-UAV systems. It combines edge caching, causal reasoning, and hybrid verification to check both logical security invariants and physical UAV constraints. DER~\cite{P2497} decomposes LLM-driven UAV control into a Decomposer, an Executor, and a Replanner. The Executor constrains UAV tool calls through sandboxed invocations and runtime guards, while the Replanner triggers recursive replanning according to mission state and anomaly detection.

\textbf{Optimization-mediated multi-agent coordination.} Optimization-mediated multi-agent coordination combines LLM reasoning, human feedback, and optimization solvers for hazardous multi-agent environments. Its core idea is to restrict the LLM to high-level decisions, such as task reconfiguration, objective-weight adjustment, and human-feedback interpretation. Low-level actions are instead solved by an optimizer under safety constraints, reducing execution risks caused by directly using LLM outputs as control commands.

Hierarchical LLMs In-the-Loop Optimization~\cite{P4762} formulates multi-robot target tracking under sensing and communication hazards as a bi-level optimization problem. It uses a low-frequency Task LLM for strategic reconfiguration, a high-frequency Action LLM to adjust weights for tracking, energy, and safety, and an optimization solver to output actions. A human supervisor can provide natural-language feedback on unmodeled hazards or performance problems.

\textbf{Privacy and HRI governance.} Privacy and HRI governance mechanisms concern user data, trust calibration, and interaction risks during long-term deployment. They aim to make the system more controllable and interpretable while reducing privacy exposure and trust mismatch.

MALLM~\cite{P4343} uses personalized agents and local processing in multi-user apartment environments to handle preference conflicts and privacy protection in shared spaces. The FM-driven HRI risk framework~\cite{P008} maps foundation-model roles in HRI to the Interaction Loop and the Robotic System, and analyzes risks across multiple dimensions. The former provides a privacy-governance practice for long-term edge deployment, while the latter supports identification of human-robot interaction risks.

\subsubsection{Evaluation}

Evaluation at this stage concerns post-deployment system performance, including communication security, interaction trustworthiness, and long-term operation. Existing evaluations can be divided into deployment-architecture evaluation and interaction-trustworthiness evaluation.

Some work treats the deployment architecture itself as the evaluation object. It focuses on system-level metrics such as communication latency, provenance integrity, and data sovereignty, and analyzes a deployment chain formed by MQTT, Home Assistant, and Android edge nodes in an edge-local agent swarm~\cite{P148}. This asks whether the architecture provides security properties such as message-source verification, audit integrity, and data sovereignty.

An interactive trustworthiness benchmark for MLLM-based embodied agents further evaluates embodied interaction across multiple trustworthiness dimensions through 150 interactive tasks~\cite{P5485}.

\subsection{Cross-lifecycle Risk Propagation, Defense Composition, and Future Directions}
\label{sec:cross-lifecycle-risk-propagation}

In real environments, some security and privacy risks in large-model-driven embodied agents propagate across stages rather than remaining at the stage where they enter the system. We therefore further analyze how risks propagate across stages, and how defense mechanisms can be composed across different levels.

\subsubsection{Cross-lifecycle Risk Propagation}

The core feature of cross-lifecycle risk propagation is the misalignment among attack entry point, consequence realization point, and defense intervention point. An attack may enter at one stage but remain latent until later components transform it into downstream effects. This misalignment arises from multi-level dependencies in embodied closed loops: sensor inputs shape semantic interpretation, semantic interpretation affects task planning, task planning affects action generation, and action execution may further influence later decisions through environmental feedback and deployment records.

\textbf{Propagation from training-time risks to execution.} Data poisoning, fine-tuning backdoors, or supply-chain compromise may remain latent during model construction and alter action generation only when later triggers appear. BackdoorVLA~\cite{P1436} injects visual or textual triggers into training samples and binds them to attacker-specified long-horizon action trajectories, causing targeted execution at inference time. State Backdoor~\cite{P1422} shifts the trigger from an externally visible object to the robot's proprioceptive state, enabling activation through the initial state. DropVLA~\cite{P1441} and SILENTDRIFT~\cite{P1424} show that payloads can enter action windows, low-level primitives, or trajectory structures, appearing as object dropping, grasp failure, or smooth trajectory drift. GoBA~\cite{P5510} and BEAT~\cite{P5539} further demonstrate physical-object or visual-scene triggers, allowing training-time poisoning to reach execution through later environmental inputs.

\textbf{Propagation from input risks to command and action interfaces.} L2 attacks enter through sensor signals, physical text, or language prompts, and their impact depends on whether these inputs are later treated as valid information for decision-making. Phantom Menace~\cite{P1433} shows a path from physical sensor manipulation to action execution: poisoned camera and microphone inputs enter the VLA sensor-to-action pipeline and produce abnormal real-robot behavior. CHAI~\cite{P5541} shows a path from environmental text to the command layer, where natural-language text embedded in physical scenes is interpreted by the LVLM as a control instruction and affects UAVs, vehicles, or tracking systems. Textual action attacks against robotic VLAs further show that modifying only the initial rollout prompt can induce fixed actions or action sequences that remain effective over later rollout steps~\cite{P5507}.

\textbf{Propagation from planning-layer risks to physical execution.} L3 risks often appear as plans, policies, or API calls that are both harmful and executable. POEX~\cite{P5498} defines the jailbreak target as harmful yet executable robot policies, emphasizing that the attack output must bypass safety alignment while remaining executable by the robot. BADROBOT~\cite{P5483} uses contextual jailbreak, safety misalignment, and conceptual deception in voice-based interactions to induce embodied LLM frameworks to generate physically malicious behavior. ROBOPAIR~\cite{P5506} further adds robot-specific system prompts and a syntax checker, so that attack prompts produce executable instructions satisfying robotic API format requirements.

\textbf{Feedback from deployment-layer risks to planning and execution.} Communication and collaboration mechanisms at L5 can also become risk entry points and directly affect planning, control, or collective behavior. The MITM attack against an LLM-enabled vacuum robot edits robot--LLM messages and turns response manipulation into hardware-output manipulation~\cite{P5505}. InfectBot~\cite{P1275} demonstrates lateral propagation in multi-robot systems: the attacker manipulates one entry robot and then uses peer communication to spread unsafe intent to the robot group.

\textbf{Cross-boundary system structures amplify propagation paths.} DFD threat modeling~\cite{P1279} indicates that system modules are connected by cross-boundary data flows rather than operating independently. A local vulnerability at one stage may therefore propagate downstream along the system chain and eventually affect planning decisions or physical execution outcomes.

\subsubsection{Cross-lifecycle Defense Composition}

Cross-lifecycle risk propagation shows that defenses for large-model-driven embodied agents cannot rely on a single checkpoint. A more suitable strategy is to build a cross-lifecycle composition framework in which defenses impose continuous constraints at critical points and progressively weaken risk propagation. Based on the studies discussed above, we propose a five-layer defense composition framework: provenance and update control, input and message admission, semantic safety contract, runtime constraint enforcement and recovery, and audit and governance loop.

\textbf{Provenance and update control} targets model construction and updating. Its goal is to establish source trustworthiness and update traceability before risks enter the system. Training data and model components should have provenance records and version-management mechanisms. Although this layer cannot fully eliminate training-time backdoors or supply-chain compromise, it supports later risk localization, model rollback, and accountability tracing.

\textbf{Input and message admission} targets external information entering the operational closed loop, including inputs from users, environments, and system interactions. Its core task is not merely filtering ``harmful text'', but judging input source, semantic role, and execution authority. Text observed in the environment should not be automatically treated as a user command, and messages from peer agents or cloud services should not enter planning or control chains without identity authentication and integrity verification. This layer covers both L2 multimodal inputs and L5 communication messages, preventing untrusted inputs from being treated downstream as valid task conditions.

\textbf{Semantic safety contract} operates at the semantic reasoning and task-planning stage. It translates task goals, safety requirements, and environmental constraints into checkable safety-constraint representations for action-generation and control-execution modules. SafeGate~\cite{P1280}, Safety Chip~\cite{P5242}, and cross-layer sequence supervision~\cite{P5524} illustrate paths from natural-language tasks to safety contracts, LTL constraints, and cross-layer supervisors. The key challenge is constraint coverage and environmental grounding. If hazardous states, object relations, and action side effects are not correctly modeled, formal constraints will still miss risks.

\textbf{Runtime constraint enforcement and recovery} acts on action generation, control execution, and environmental feedback. It blocks behaviors that pass upstream checks but may still cause physical risks. Language-conditioned safety filtering~\cite{P1311} and SafeDec~\cite{P1689} show that safety constraints can enter control filtering or action decoding rather than remain only at the planning layer. Runtime defense should impose constraints during action generation or control execution and trigger safe stop or replanning after abnormality detection. It must also define recovery paths; otherwise, the system may detect an anomaly but fail to handle it safely.

\textbf{Audit and governance loop} targets long-term deployment and post-incident governance. Its goal is to extend one-time safety checks into continuous system governance. Deployed systems should record key operational information and abnormality-handling processes for incident reconstruction and accountability tracing. DER~\cite{P2497}, Decentralized LLM oracle~\cite{P1324}, and FlyAdapt~\cite{P2230} provide partial mechanisms for runtime recovery, task-sequence verification, and access control. However, these methods remain scenario-specific. A general governance layer requires unified identity and permission management, together with auditable operational mechanisms, to maintain stability during long-term operation.

Existing studies have provided local mechanisms at these layers, but they have not yet formed a systematically validated end-to-end defense stack. Future research should further evaluate how these layers transmit constraints, share states, handle conflicts, and report residual risks.

\section{Gaps and Future Directions}
\label{sec:gaps-future-directions}

Existing studies have covered multiple stages in the lifecycle of large-model-driven embodied agents, but the field remains fragmented across stages, uneven in evidentiary depth, and insufficient in system-level governance. Future research therefore needs to move beyond point-wise attacks and defenses toward lifecycle-wide security governance.

\textbf{Model construction requires stronger provenance and supply-chain auditing.} Existing L1 defensive research remains limited, and the field still lacks toolchains for systematically detecting latent backdoors and component contamination. Future work should establish unified provenance and auditing mechanisms to support traceability and verification across data sources, model components, and external services. Moreover, training membership leakage and capability removal still lack strong guarantees~\cite{P1322,P5512}, making certified unlearning, membership auditing, and privacy-preserving continual learning important directions.

\textbf{Input-stage research must address compound cross-modal risks.} Most L2 studies isolate one channel, although embodied agents integrate diverse multimodal inputs. Their joint effects and propagation into planning or execution remain underexplored. Future work should model compound attacks, identify cross-modal amplification, and develop defenses that authenticate input sources and protect multiple input classes without relying on channel-specific assumptions.

\textbf{Planning-stage research needs to improve both task competence and safety awareness.} Existing studies indicate a clear disconnect between planning capability and safety awareness, while attacks have moved beyond text-level manipulation to executable plans and policies. Text-level rejection mechanisms alone are insufficient for planning-phase risks. Future work should incorporate executability and permission constraints into planning-security analysis and develop safety-aware planning mechanisms that connect natural-language risk descriptions with environmental states.

\textbf{Execution-stage research needs to move from point-wise control filtering to closed-loop recovery.} Existing execution-layer defenses mainly rely on control filtering and anomaly detection, many of which still assume reliable perception, approximate dynamics, or predefined constraints. Future research should advance execution safety from isolated control mechanisms toward closed-loop recovery frameworks, enabling systems to respond to and recover from anomalies.

\textbf{Deployment-stage research requires systematic governance and long-term privacy evaluation.} Communication links, collaboration protocols, and human-agent trust mechanisms may all become attack surfaces. Existing defenses often depend on specific platforms or tasks and have not formed a unified cross-system security framework. Long-term privacy evaluation also remains insufficient, lacking systematic quantification of risks that emerge over prolonged use. Future work should develop integrated frameworks for continuously tracking privacy leakage, trust dynamics, and system accountability.

\textbf{Cross-stage evaluation needs to form a unified chain of evidence.} Current benchmarks cover several key security directions, but evidence levels remain insufficiently standardized and are difficult to compare or substitute directly. Future work should construct cross-stage evidence chains that record and compare how risks arise, propagate, and manifest, enabling more reliable assessment of attack and defense mechanisms within a complete embodied closed loop.

\section{Conclusion}
\label{sec:conclusion}

This survey reviewed security and privacy issues in large-model-driven embodied agents. We first defined the survey object by limiting the scope to agent in which large-model capabilities substantially participate in embodied closed-loop operation. We also extended the analysis boundary from model outputs to the complete embodied-agent system. On this basis, we proposed a five-stage lifecycle framework and organized related research work under these stages. This framework integrates previously scattered studies into a unified system process, facilitating the analysis of attack mechanisms, defense mechanisms, evaluation methods, and research gaps throughout the full system lifecycle.

The stage-wise analysis leads to three main observations. First, attack research has expanded from model outputs to training data, sensor inputs, semantic planning, action execution, and deployment communication. Its consequences have also extended from incorrect answers to trajectory deviation, task failure, control tampering, and privacy leakage. Second, defense methods now cover alignment, input moderation, formal verification, runtime control, and governance. However, many methods still depend on specific environments or assumptions, and composable system-level solutions remain underdeveloped. Third, evaluation has moved beyond textual safety toward planning, execution, and real-robot validation, but different evidence levels still lack a unified standard.

Overall, research in this area needs to move from pointwise protection toward systematic governance, and support risk assessment and accountability in real deployment with clear evidence.

\bibliographystyle{ACM-Reference-Format}
\bibliography{references}










\end{document}